\documentclass[a4paper,fleqn,usenatbib]{mnras}

\usepackage{newtxtext,newtxmath}

\usepackage[T1]{fontenc}
\usepackage{ae,aecompl}

\usepackage{graphicx}	
\usepackage{amsmath}	
\usepackage{amssymb}	
\usepackage{xspace}
\usepackage{subfig}
\usepackage{enumitem}

\newcommand{\icg}{{Institute of Cosmology \& Gravitation, University of Portsmouth,
Dennis Sciama Building, Burnaby Road, Portsmouth PO1 3FX, UK}}
\newcommand{\uom}{{School of Physics, University of Melbourne, Parkville, VIC 3010, Australia}}
\newcommand{\qmu}{{School of Physics and Astronomy, Queen Mary University of London, Mile End Road, London E1 4NS, UK}}
\newcommand{\uwc}{{Department of Physics \& Astronomy, University of the Western Cape, Cape Town 7535, South Africa}}

\newcommand{\hi}{H\textsc{i}\ }
\newcommand{\hiindex}{\textrm{H\textsc{i}}}

\newcommand{\size}[2]{{\fontsize{#1}{0}\selectfont#2}}

\newcommand{\fastica}{\size{7.5}{FASTICA }}
\newcommand{\euclid}{\textit{Euclid}\xspace}

\title[Impact of Foregrounds on HI $\times$ Optical Cross-Correlations]{Impact of Foregrounds on HI Intensity Mapping Cross-Correlations with Optical Surveys}

\author[S.Cunnington et al.]{Steven Cunnington$^{1}$\thanks{E-mail: steve.cunnington@port.ac.uk}, Laura Wolz$^{2}$, Alkistis Pourtsidou$^{3,4}$, David Bacon$^{1}$\\
$^{1}$\icg\\
$^{2}$\uom\\
$^{3}$\qmu\\
$^{4}$\uwc\\
}

\date{Accepted XXX. Received YYY; in original form ZZZ}

\pubyear{2018}

\begin{document}
\label{firstpage}
\pagerange{\pageref{firstpage}--\pageref{lastpage}}
\maketitle


\begin{abstract}
The future of precision cosmology could benefit from cross-correlations between intensity maps of unresolved neutral hydrogen (\hiindex) and more conventional optical galaxy surveys. 
A major challenge that needs to be overcome is removing the 21cm foreground emission 
that contaminates the cosmological \hi signal. Using $N$-body simulations we simulate \hi intensity maps and optical catalogues which share the same underlying cosmology. Adding simulated foreground contamination and using state-of-the-art reconstruction techniques we investigate the impacts that 21cm foregrounds and other systematics have on these cross-correlations. We find that the impact a \fastica 21cm foreground clean has on the cross-correlations with spectroscopic optical surveys with well-constrained redshifts is minimal. However, problems arise when photometric surveys are considered: we find that a redshift uncertainty $\sigma_z \geq 0.04$ causes significant degradation in the cross power spectrum signal. We diagnose the main root of these problems, which relates to arbitrary amplitude changes along the line-of-sight in the intensity maps caused by the foreground clean and suggest solutions which should be applicable to real data. These solutions involve a reconstruction of the line-of-sight temperature means using the available overlapping optical data along with an artificial extension to the \hi data through redshift to address edge effects. We then put these solutions through a further test in a mock experiment that uses a clustering-based redshift estimation technique to constrain the photometric redshifts of the optical sample. We find that with our suggested reconstruction, cross-correlations can be utilized to make an accurate prediction of the optical redshift distribution.
\end{abstract}

\begin{keywords}large-scale structure of Universe  -- distances and redshifts -- cosmology: observations -- techniques: spectroscopic -- photometric -- radio lines: galaxies
\end{keywords}



\section{Introduction}

Conventional optical galaxy surveys map large-scale cosmic structure by using resolved galaxies as tracers of the underlying dark matter field, which dominates the overall matter density. Their distribution contains information about the expansion history and the growth of cosmic structure. Using cosmological probes such as baryon acoustic oscillations we can measure the Universe's expansion history and constrain dark energy \citep{Percival:2001hw}. Measuring the large-scale distribution of matter also reveals information on the primordial state of the Universe \citep{Slosar:2008hx} which has the potential to constrain models of the early Universe. Furthermore, probing structure on cosmic horizon scales will test General Relativity and indicate if modifications to this theory of gravity are required.

Methods involving detection of galaxies to trace large-scale structure are reliable providing that the galaxy samples obtained by a survey have a sufficient number density. If not, the measurements will suffer from significant statistical errors due to Poisson shot noise. Obtaining a large number of resolved galaxies with precise redshifts is expensive; spectroscopic redshifts with a redshift uncertainty $\sigma_z \sim 0.001$ rely on long integration times making this a slow process. Photometric redshifts offer a less precise alternative but can be obtained much more quickly allowing dense catalogues of galaxies to be built \citep{Bolzonella:2000js, FernandezSoto:2000vu}. It is for this reason that future stage-IV surveys such as \euclid\footnote{\href{www.euclid-ec.org/}{www.euclid-ec.org/}} \citep{Amendola:2016saw} will heavily rely on photometric redshifts, and the Large Synoptic Survey Telescope (LSST)\footnote{\href{www.lsst.org/}{www.lsst.org/}} \citep{Marshall:2017wph} will be entirely reliant on them.

As an alternative, radio intensity mapping techniques, which do not rely on resolving individual sources, offer the prospect of more complete tracer maps with the redshift precision of a spectroscopic survey. In complete contrast to optical surveys, intensity mapping provides excellent constraints along the radial line-of-sight but poor angular resolution. This complementarity, together with the fact that cross-correlations are expected to alleviate survey-specific systematic effects, makes synergies between intensity mapping and optical galaxy surveys mutually beneficial. 

Arguably the most appealing source of emission for cosmology with intensity mapping comes from the neutral hydrogen (\hiindex) gas residing within galaxies.  The signature of \hi that we aim to detect comes from the hyperfine transition of hydrogen's single electron. When the electron drops to a lower energy state it emits a photon with a rest frequency of $1420$ MHz, or 21cm of wavelength, hence the synonymous name $21$cm intensity mapping. The observed frequency of these emitted photons places the signal in the radio part of the electromagnetic spectrum. Therefore radio dishes are the conventional choice of receiver for detecting these photons at low redshifts of $z<3$. First detections using the \hi  intensity mapping technique have already been achieved in cross-correlation with optical galaxies in \citet{PenIMfirstdetection}, \citet{GBTHIdetection1} and \citet{ParkesIMxOptDetection}.

The most prominent example of a next generation radio observatory is the Square Kilometre Array (SKA)\footnote{\href{www.skatelescope.org/}{www.skatelescope.org/}} \citep{Bacon:2018dui}. The mid-frequency instrument, SKA1-MID (where $1$ stands for Phase $1$), will be an array of 197 dish receivers that can operate in interferometer and single-dish mode. The low frequency instrument, SKA1-LOW, will probe the high redshift Universe, targeting the Epoch of Reionisation. As with any interferometer, it is the largest separation (or baseline) which determines the resolution of the instrument; hence baselines of up to $150$ km are proposed to maximize resolution. Conversely, it is the smallest baselines between receivers which determines the largest scales that can be probed. The SKA1-MID instrument aims to perform a wide ($\sim 20,000\ \text{deg}^2$) \hi intensity mapping survey in single-dish mode. This compromises angular resolution but probes the large scales needed for cosmology.

\begin{table}
	\centering
	\begin{tabular}{cccccc}
    	\hline
		21cm IM Survey & & Photo-$z$ Survey & $f_\text{sky}$ & $z_\text{min}$ & $z_\text{max}$ \\
        \hline
        MeerKAT & $\times$ & DES & 0.1 & 0 & 1.45\\
        TIANLAI & $\times$ & DECaLS & 0.15 & 0 & 1.5\\
		SKA1-MID & $\times$ & \euclid & 0.2 & 0.35 & 2\\
		SKA1-MID & $\times$ & LSST & 0.4 & 0.35 & 3\\
		HIRAX & $\times$ & \euclid & 0.2 & 0.8 & 2\\
		HIRAX & $\times$ & LSST & 0.5 & 0.8 & 2\\
		CHIME & $\times$ & \euclid & 0.35 & 0.8 & 2\\
		CHIME & $\times$ & LSST & 0.5 & 0.8 & 2.5\\
		\hline
	\end{tabular}
    \caption{Examples of cross-correlation opportunities between 21cm intensity mapping surveys and optical photometric redshift surveys, with (approximate) estimates for their sky and redshift overlap. $f_\text{sky}$ refers to the fraction of full sky for which these surveys can overlap. $z_\text{min}$ and $z_\text{max}$ represent the common redshift overlap range.}
    \label{CrosssSurveys}
\end{table}

The redshifted 21cm line signal from \hi benefits from being particularly isolated in frequency, and there are few examples of spectral lines that could lead to potential line confusion, making \hi intensity mapping particularly robust for redshift experiments. However, a major challenge for \hi intensity mapping comes from foreground emission (e.g. synchrotron radiation), which can be orders of magnitude larger than the cosmological signal. Foregrounds are spectrally smooth signals which emit in the same range as the redshifted \hi. 
Blind foreground removal techniques, which require no prior knowledge or templates of the foregrounds, can be used to exploit the smooth form of most foreground signals along the line-of-sight to isolate and remove them.

In this work we investigate how foreground removal can impact important cosmological measurements. Several studies have investigated how foreground removal can be carried out without detrimental impact on the \hi auto-correlation power spectrum recovery \citep{Wolz:2013wna, Shaw:2013wza, Alonso:2014dhk}. In this work we aim to place particular emphasis on the foreground removal's impact on cross-correlation measurements with optical galaxy surveys. Examples of some future optical-21cm cross-correlation possibilities are outlined in Table \ref{CrosssSurveys}. In order to investigate the impact of foregrounds on cross-correlations, we utilize mock galaxy catalogues built from $N$-body simulations of dark matter particles. This approach allows for both optical and \hi intensity map data to share the same underlying simulated cosmology, with realistic parameters (such as number density of galaxies) corresponding to the specifications of current and forthcoming surveys.

The plan of the paper is as follows:
In Section \ref{CosmologicalSig} we describe how we simulate the cosmological signals, both our resolved optical galaxy number density maps and the overlapping \hi intensity maps. Section \ref{FG} explains how we simulate the 21cm foregrounds, which are then added into our \hi cosmological signal to contaminate the intensity maps. Section \ref{FGremoval} then explains the processes used for removing these foregrounds and details the \fastica approach that we opt to use on our simulations. In Section \ref{HIxOpticalSec} we analyze our results and demonstrate what impact foreground cleaning can have on a cross-correlation power spectrum. In Section \ref{ClusteringzSec} we extend these findings and apply them to a practical experiment which utilizes these cross-correlations to constrain photometric redshifts using \hi intensity maps. We conclude and discuss in Section \ref{conclusion}.

\section{Cosmological Signals \& Their Simulation}\label{CosmologicalSig}

In order to probe the large-scale cosmic structure and map the matter over-density $\delta$, we rely on luminous sources which trace the underlying matter density. In optical galaxy redshift surveys we use number density fields $n_\text{g}(\vec{\theta},z)$ where resolved galaxies can be counted in voxels (3-dimensional pixels) at angular position $\vec{\theta}$ with a redshift $z$ which is used for defining the line-of-sight (LoS) distance. We can then calculate the over-density of galaxies $\delta_\text{g}$, which we assume is a linearly biased tracer of the matter over-density $\delta$, i.e.
\begin{equation}
    \delta_\text{g}(\vec{\theta},z) \equiv \frac{n_\text{g}(\vec{\theta},z) - \overline{n}_\text{g}(z)}{\overline{n}_\text{g}(z)} = b_\text{g}(z)\delta (\vec{\theta},z) \, ,
\end{equation}
where a barred quantity represents a mean average and $b_\text{g}$ is the (linear) galaxy bias.

For \hi intensity maps there are no resolved luminous sources, only combined brightness temperatures in a given voxel. 
Assuming that \hi is also a biased tracer of the underlying matter density we can write
\begin{equation}
    \delta_\hiindex(\vec{\theta},z) \equiv \frac{T_\hiindex(\vec{\theta},z) - \overline{T}_\hiindex(z)}{\overline{T}_\hiindex(z)} = b_\hiindex(z)\delta (\vec{\theta},z) \, ,
\end{equation}
where the linear bias factor is now $b_\hiindex$. Note that the mean brightness temperature $\overline{T}_\hiindex$ is also an unknown quantity, degenerate with $b_\hiindex$. Since the observable is a temperature fluctuation, it is customary to work with temperature fluctuation maps where
\begin{equation}
    \delta T_\hiindex(\vec{\theta},z) \equiv T_\hiindex(\vec{\theta},z) - \overline{T}_\hiindex(z) = \overline{T}_\hiindex(z)b_\hiindex(z)\delta (\vec{\theta},z).
\end{equation}
It is these quantities, $\delta_\text{g}$ and $\delta T_\hiindex$, which can be used to make cosmological measurements e.g. auto-power spectra $P_\text{gg} \sim  \langle |\Tilde{\delta}_\text{g}|^2 \rangle$ or cross-power spectra $P_\text{g,\hiindex} \sim \langle \text{Re}\{ |\Tilde{\delta}_\text{g}\Tilde{\delta} T^*_\hiindex|\} \rangle$. Here the tilde notation $\Tilde{\delta}$ represents the Fourier transform of the matter over-density. 

An important measurement in cosmology, and one we heavily focus on in this work, is the angular clustering of a matter density tracer. In order to apply this with \hi intensity maps, we measure the angular power spectrum by decomposing the temperature fluctuations into spherical harmonics $Y^m_\ell(\textbf{\^{n}})$:
\begin{equation}
    \delta T_\hiindex(\textbf{\^{n}},\nu) = \sum_{\ell=0}^\infty \sum_{m=-\ell}^{m=\ell} a_{\ell m}(\nu)Y^m_\ell(\textbf{\^{n}}) \, .
\end{equation}
The harmonic coefficients $a_{\ell m}(\nu)$ describe the amplitudes of the fluctuations in spherical harmonic space; we can then define the angular power spectrum between tracers $X$ and $Y$ as
\begin{equation}\label{ClPowerSpec}
    C^{XY}_\ell(\nu_1,\nu_2) = \langle a^X_{\ell m}(\nu_1) a^{Y*}_{\ell m}(\nu_2) \rangle \, .
\end{equation}
Consideration must also be given to data that does not cover the full sky and instead comes from only the footprint covered by the survey. The simulations we use will have partial sky coverage and therefore emulate this problem. This has consequences for the power spectrum and results in correlated multipoles which bias the measurement. In this work we are not particularly interested in making precise comparisons of a measured power spectrum to say one predicted by a $\Lambda$CDM model. Instead we are interested in the comparison of a power spectrum free of 21cm foregrounds to one contaminated by them, which should both be biased by cut skies in the same way. However, to ensure an accurate treatment of the cut skies we will use the pseudo-$C_\ell$ method of angular power spectrum measurement \citep{Wandelt:1998qd,Wandelt:2000av} and use the unified pseudo-$C_\ell$ framework \texttt{NaMaster}\footnote{\href{https://github.com/LSSTDESC/NaMaster}{https://github.com/LSSTDESC/NaMaster}} \citep{Alonso:2018jzx} and its \texttt{python} wrapper \texttt{pymaster}.

If the tracer fields are Gaussian, the power spectrum \eqref{ClPowerSpec} is a complete statistical representation of the fields. The power spectrum can either represent the \hi intensity map auto-correlation where $X=Y=\hiindex$, or the cross-correlation with the optical galaxies where $X=\text{g}$ and $Y=\hiindex$. Hence, in order to use simulations to study the impact 21cm foregrounds can have on cross-correlation cosmological measurements such as $C_\ell^{\text{g},\hiindex}$, we require a simulation which includes \hi emission and resolved optical galaxies.

In many 21cm studies it is sufficient to simulate wide continuous intensity maps through Gaussian realizations of a \hi power spectrum. However, for this work we need an optical galaxy catalogue which shares the same underlying cosmology as the \hi intensity maps, since we are looking to exploit a shared clustering signal between resolved optical galaxies and \hi emission for cross-correlated measurements. It is also preferable to have the optical galaxy simulation as a resolved catalogue of sources so that $N(z)$ distributions can be built precisely from individual galaxy redshifts. We can then choose to degrade the redshift accuracy in order to emulate a photometric imaging survey.

In order to achieve this, we use existing $N$-body galaxy simulations and exploit certain components of them, e.g. \hi mass or halo mass to simulate \hi brightness temperatures which we can build intensity maps from. Utilizing $N$-body simulations also allows for a more robust representation of a survey catalogue than Gaussian realized signals. With this in mind we ideally require a catalogue which has the following features:

\begin{itemize}[leftmargin=*]
\item low halo-mass resolution ($\approx 10^9 h^{-1}M_\odot$) so that intensity maps include integrated \hi emission from faint galaxies;
\item \hi information for each galaxy for simulating realistic intensity maps;
\item deep redshift and wide sky coverage ($0<z<3$, $\sim 20,000 \, {\rm deg}^2$) to allow for testing low resolutions associated with the typical beam size of a SKA-like intensity mapping experiment;
\item simulated photometry for optically resolved galaxies so that realistic cross-correlation forecasts can be made between intensity maps and photometric galaxy surveys.
\end{itemize}
A simulation including all of the above is not currently available, and is unlikely to be available in the near future. This is largely due to the fact that low halo mass resolution with sufficient galaxy number densities over large sky volumes would require $N$-body simulations that would be exceptionally computationally expensive.

In this work we therefore utilize two simulated catalogues with differing characteristics. One catalogue contains \hi signal with a low halo mass resolution and simulated \hi masses for every galaxy. The other is a more conventional optical survey catalogue with simulated photometry but which is not as resolved in halo mass. We will now describe the two catalogues in detail.

\begin{itemize}[leftmargin=*]
\item \bf{GAEA Simulation\footnote{\href{http://astrosims.flatironinstitute.org/gaea}{http://astrosims.flatironinstitute.org/gaea}}}
\end{itemize}
We make use of the GAEA semi-analytic model \citep{GAEA, GAEA2, GAEA3}. The catalogue was built using the Millennium Simulation \citep{SpringelMillennium}, which is a cosmological $N$-body simulation that used $N = 2160^3$ particles of mass
$m_\text{p} = 8.6 \times 10^8 h^{-1}M_\odot$ within a comoving box of size $500^3$ (Mpc/$h)^3$ with a cosmology consistent with WMAP1 data. In particular, the values of the adopted cosmological parameters are: $\Omega_\text{b} = 0.045,\ \Omega_\text{m} = 0.25,\ \Omega_\Lambda = 0.75,\
H_0 = 100h \, \text{Mpc}^{-1}\text{km s}^{-1},\ h = 0.73,\ \sigma_8 = 0.9$ and $n_s = 1$. The GAEA catalogue is built replicating this same box, but selecting galaxies from the nearest snapshot corresponding to its co-moving distance from the observer.

GAEA used an algorithm to identify halos which allowed for a halo mass resolution of $1.7 \times 10^{10}M_\odot h^{-1}$ which resulted in just over $2\times 10^8$ galaxies with a continuous sky coverage $f_\text{sky}=0.5$. Redshifts are limited to $0<z<0.5$ which means we will only be able to study cross-correlations within this small range, but this should still allow for multiple redshift/frequency bins given the completeness within this range. GAEA also includes simulated \hi masses for all its galaxies, which can be used to generate realistic \hi brightness temperatures. We discuss this further in Section \ref{HIIMSec}.

\begin{itemize}[leftmargin=*]
\item \bf{MICE Simulation\footnote{\href{http://maia.ice.cat/mice/}{http://maia.ice.cat/mice/}}}
\end{itemize}
We also make use of the MICECATv2.0 simulation released as part of the MICE-Grand Challenge Galaxy and Halo Light-cone Catalogue \citep{MICE1,MICE2,MICE3,MICE4,MICE5}, which is a cosmological $N$-body dark matter only simulation containing 4096$^3$ dark-matter particles of mass $m_\text{p}=2.93 \times 10^{10} h^{-1} M_\odot$ in a box-size of $3072^3$ (Mpc/$h)^3$. They resolved halos down to a few $10^{11}M_\odot h^{-1}$ and used a hybrid Halo Occupation Distribution (HOD) and Halo Abundance
Matching (HAM) technique for galaxy modelling resulting in just under $5 \times 10^8$ galaxies. The simulation's sky footprint is $90 \times 90$ deg$^2$ filling an octant of sky ($f_\text{sky}=0.125$) up to a redshift $z=1.4$. The assumed cosmology is a flat concordance $\Lambda$CDM model
with $\Omega_\text{m} = 0.25$, $\Omega_\Lambda = 0.75$, $\Omega_\text{b} = 0.044$, $n_\text{s} = 0.95$, $\sigma_8 = 0.8$ and $h = 0.7$ consistent with  WMAP 5-year data.

\begin{table}
	\centering
	\begin{tabular}{cccccc} 
		\hline
		Catalogue & Box Volume & $m_\text{p}$ & N$_\text{gal}$ & $f_\text{sky}$ & $z_\text{max}$ \\
		& [$(\text{Mpc}/h)^3$] & [$M_\odot / h$] & & & \\
        \hline
		GAEA & $500^3$ & 8.6 $\times 10^8$ & $201 \times 10^6$ & 0.5 & 0.5\\
		MICE & $3072^3$ & 2.9 $\times 10^{10}$ & $497 \times 10^6$ & 0.125 & 1.4\\
		\hline
	\end{tabular}
    \caption{Summary of the two different mock galaxy catalogues we will be using. Both are built from $N$-body simulations for which we provide the box size and particle mass $m_\text{p}$.}
    \label{SimSummary}
\end{table}

Since the MICE catalogue does not have simulated \hi masses for each galaxy, we must derive our own. We therefore take each central galaxy's halo mass as simulated by MICE and convert this into a predicted \hi mass by following the redshift dependent prescription laid out in \citet{Padmanabhan:2016tcc}
\begin{equation}\label{PadmanabhanM_HI}
	M_\hiindex = 2N_{1}M\bigg[ \bigg(\frac{M}{M_{1}}\bigg)^{-b_{1}} + \bigg(\frac{M}{M_{1}}\bigg)^{y_{1}} \bigg]^{-1} \, ,
\end{equation}
where $M$ is the galaxy's halo mass; $M_1$, $N_1$, $b_1$ and $y_1$ are all free parameters with redshift dependence tuned to provide a best fit; we refer the reader to \citet{Padmanabhan:2016tcc} for details. Each central galaxy then has a \hi mass from which we can generate a \hi brightness temperature signal. While this prescription would not be ideal for small scale studies of \hi distribution, since we are assuming that all \hi lies within central galaxies, it suits our purposes because we will be smoothing out any small scale imprecisions when we simulate the effect of an intensity mapping beam.
\\
\\
\noindent From these catalogues, which we summarize in Table \ref{SimSummary}, we will produce both our \hi intensity maps (Section \ref{HIIMSec}) and our detected optical galaxy catalogue (Section \ref{PhotozcatSec}), which will share the same underlying dark-matter distribution. It is this shared clustering signal which we will look to utilize in our cross-correlation tests. We emphasize once more that we use these two separate $N$-body simulations since each one has unique advantages. For example the semi-analytical GAEA has replication of the particle box sample which delivers larger sky sizes and also has \hi masses for each galaxy at lower mass resolution. Both of these features contribute to delivering more robust simulations of large-beam \hi intensity maps. In contrast MICE uses a HOD/HAM approach over a larger box size, so is arguably more realistic in its cosmological signal in that no replication is required, but perhaps less realistic in that it distributes synthetic galaxies into simulated halos using a statistical approach rather than simulating baryonic process to drive galaxy evolution, as performed in semi-analytic models. MICE also includes some simulated photometric redshifts which we can utilize for forecasting the impacts of \hi foregrounds in cross-correlations with a photometric survey.

\subsection{HI Intensity Map Simulation}\label{HIIMSec}

We aim to express our \hi intensity map data $T_\hiindex$ in the form of a brightness temperature with two angular dimensions ($\theta_\text{ra}$ and $\theta_\text{dec}$, jointly represented by $\vec{\theta}$) and a radial dimension, the redshift $z$. To do this we follow the same recipe laid out in \citet{Cunnington:2018zxg} which we repeat here for completeness.

To construct $T_\hiindex$ we start with the \hi mass $M_\hiindex$ of each galaxy, which is given in the GAEA catalogue and generated using halo masses and Equation (\ref{PadmanabhanM_HI}) for MICE. We then place our galaxies into a data cube with coordinates ($\theta_\text{ra}, \theta_\text{dec}, z$) by binning each galaxy's \hi mass into its relevant voxel so we end up with a gridded \hi mass map $M_\hiindex(\vec{\theta},z_c)$.

We can then convert this into an intensity field for a frequency width of $\delta \nu$ subtending a solid angle $\delta \Omega$ (which is effectively our pixel size)
\begin{equation}
    I_\hiindex(\vec{\theta},z) = \frac{3h_\text{P}A_{12}}{16\pi m_\text{h}}\frac{1}{\left[(1+z)\chi(z)\right]^2}\frac{M_\hiindex(\vec{\theta},z)}{\delta \nu \delta \Omega}\nu_{21} \, ,
\end{equation}
where $h_\text{P}$ is the Planck constant, $A_{12}$ the Einstein coefficient which quantifies the rate of spontaneous photon emission by the hydrogen atom, $m_\text{h}$ is the mass of the hydrogen atom, $\nu_{21}$ the rest frequency of the 21cm emission and $\chi(z)$ is the comoving distance out to redshift $z$ (we will assume a flat universe).

It is conventional in radio astronomy, in particular intensity mapping, to use brightness temperature which can be defined as the flux density per unit solid angle of a source measured in units of equivalent black body temperature. Hence, our intensity $I_\hiindex(\vec{\theta},z)$ can be written in terms of a black-body temperature in the Rayleigh-Jeans approximation $T=Ic^2/(2k_\text{b}\nu^2)$ where $k_\text{b}$ is the Boltzmann constant. Using this we can estimate the brightness temperature at redshift $z$
\begin{equation}
    T_\hiindex(\vec{\theta},z) = \frac{3h_\text{P}c^2A_{12}}{32\pi m_\text{h}k_\text{b}\nu_{21}}\frac{1}{\left[(1+z)\chi(z)\right]^2}\frac{M_\hiindex(\vec{\theta},z)}{\delta \nu \delta \Omega} \, .
\end{equation}
For cosmology studies one aims to make measurements at different redshifts. We therefore choose to slice the intensity maps into thin tomographic redshift bins and collapse these to a 2D slice which can be auto-correlated or cross-correlated with another survey map. We will often discuss binning by frequency ($\nu$) or redshift ($z$). To clarify, these are interchangeable expressions since $z+1 = \nu_{21}/\nu_\text{obs}$. For consistency however, bin widths will always be constant in redshift. The width of each tomographic redshift bin needs to be thin enough that we can make certain thin bin assumptions, yet wide enough that we allow for sufficient structure to obtain a strong cross-correlation signal. By thin bin assumptions we are referring to cosmological quantities such as the bias, which we assume to be constant within the width of our bin ($\Delta z = 0.02, 0.05$ for GAEA and MICE respectively).

In order to ensure our \hi intensity map amplitudes are in agreement with what is theoretically predicted, we choose to rescale each redshift bin so that it agrees with a model average temperature $\overline{T}_\hiindex$. For example \citet{BattyeBINGOSingleDish} gives this average temperature as
\begin{equation}\label{TbarModelEq}
    \overline{T}_\hiindex(z) = 180\Omega_{\hiindex}(z)h\frac{(1+z)^2}{H(z)/H_0} \, {\rm mK}
\end{equation}
where $\Omega_\hiindex$ is the \hi density (abundance). In principle $\Omega_\hiindex$ can be measured using the auto-correlation \hi power spectrum with redshift space distortions, assuming a fixed fiducial cosmology \citep{GBTHIdetection1,AlkistisIMoptCMBcross}. For this work we use a fit for the \hi density \citep{Bacon:2018dui}
\begin{equation}
    \Omega_\hiindex(z) = 0.00048+0.00039z-0.000065z^2 \, .
\end{equation}
In radio \hi intensity mapping the observable signals detected by a telescope are brightness temperature fluctuations,
\begin{equation}\label{deltaTeq}
    \delta T_\hiindex(\vec{\theta},z) = T_\hiindex(\vec{\theta},z) - \overline{T}_\hiindex(z) \, .
\end{equation}
We will therefore convert all our intensity maps to these quantities.

\subsubsection{Receiver Noise}\label{noiseintro}

\begin{figure}
  	\includegraphics[width=\columnwidth]{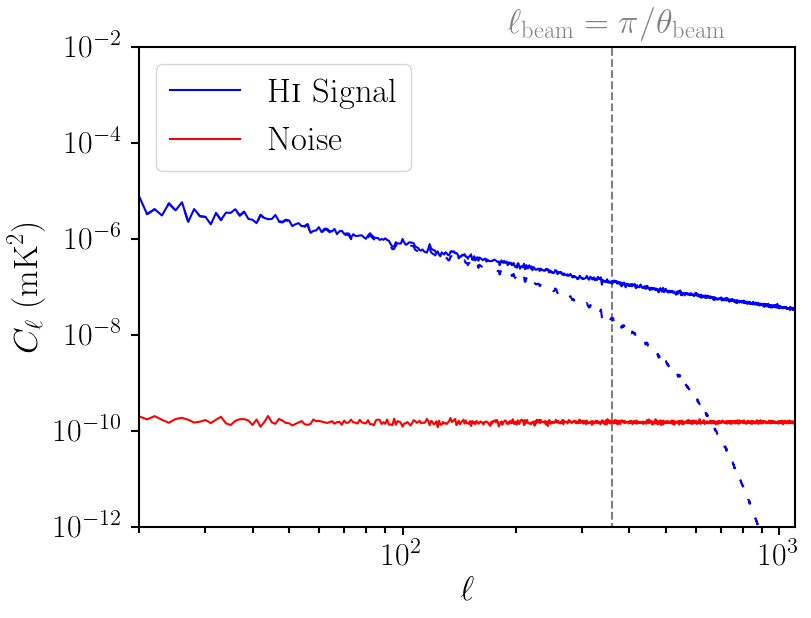}
    \caption{Angular power spectrum at a redshift of $z=0.25$ ($\nu = 1136$ MHz) for both the cosmological signal (blue solid line) for a \hi intensity map produced using the GAEA catalogue, and instrumental noise (red solid line). Also included is the effect of a $\theta_\text{beam}=0.5^\text{o}$ Gaussian convolution (blue dashed line) which shows a degradation in the cosmological \hi signal on smaller scales (high $\ell$). The grey vertical dashed line shows the angular scale of this beam. We see that instrumental noise begins to dominate at around $\ell > 700$.}
    \label{NoiseCl}
\end{figure}

As we are aiming to simulate realistic observations, we need to include the effects of instrumental (thermal) noise.
For the case of a single-dish intensity mapping experiment instrumental noise can be modelled as uncorrelated Gaussian fluctuations. Following \citet{Alonso:2014dhk} and \citet{Santos:2015gra} we add onto our observable maps a Gaussian random field with rms
\begin{equation}\label{NoiseEq}
    \sigma_\text{noise} = T_\text{sys} \sqrt{\frac{4\pi f_\text{sky}}{\Omega_\text{beam}N_\text{dish}t_\text{obs}\Delta \nu}} \, .
\end{equation}
Here $T_\text{sys}$ is the total system temperature which is the sum of the sky and receiver noise, $T_\text{sys} = T_\text{rcvr} + T_\text{sky}$ with $T_\text{rcvr} = 0.1T_\text{sky} + T_\text{inst}$ and $T_\text{sky}(\nu) \approx 60(300\text{MHz}/\nu)^{2.55} \, \text{K}$. We set $T_\text{inst}=20$ K which is representative of SKA1-MID for the redshift range $0<z<0.58$. $\Omega_\text{beam} = 1.133\theta_\text{beam}^2$ is the solid angle for the intensity mapping beam. We also assume SKA1-MID-like values for the remaining variables in the noise model with the fraction of sky $f_\text{sky}=0.41$ (representative of an SKA-LSST overlap), the number of dishes $N_\text{dish}=197$ and the total observation time $t_\text{obs}=10,000$ hours. Lastly, $\Delta \nu$ is the frequency bandwidth for a particular redshift bin. Figure \ref{NoiseCl} shows the level of this noise in relation to the cosmological \hi signal. We can see that the noise only begins to dominate when the signal has the telescope beam effects (discussed in next section) included, and this is only at small scales (high $\ell$).

A complete noise simulation would require the inclusion of red noise (or 1/$f$ noise) which originates from time correlated gain fluctuations unique to radio receivers \citep{Harper:2017gln}. Here we assume that using an appropriate scan speed ($\sim 1 \, {\rm deg s}^{-1}$), as well as component separation techniques, this noise can be removed \citep{Harper:2017gln}. 
There is also an argument to include the effects of cross-shot noise caused by \hi emitting galaxies, which provide signal in the intensity maps, also being present in the optical galaxy sample \citep{Wolz:2018svc}. We assume these additional noise effects are sub-dominant at the scales of interest and do not include them in our simulations. 

\subsubsection{Beam Resolution}\label{beamintro}

To model the low angular resolution of an intensity map, we convolve $\delta T_\hiindex$ with a telescope beam in Fourier space making use of the convolution theorem. Our telescope beam is modelled as a symmetric, two-dimensional Gaussian function with a full width half maximum of $\theta_\text{beam}$ acting only in the directions perpendicular to the LoS (as the frequency/redshift resolution is excellent). The beam size can be determined by the dimensions of the radio receiver and the frequency which is being probed \citep{AlonsoIMClusteringz}:
\begin{equation}\label{beamequation}
	\theta_\text{beam} = \frac{1.22c}{\nu D_\text{max}} \, ,
\end{equation}
where $D_\text{max}$ is the maximum baseline of the radio telescope; for a single dish receiver, $D_\text{max}$ is given by the dish diameter. The GAEA redshift range of $0<z<0.5$ would mean we are looking at beam sizes of $0.99^\text{o}<\theta_\text{beam}<1.45^\text{o}$ for our intensity maps, where we have assumed a maximum baseline  of $D_\text{max} = 15$ m which is representative of the SKA1-MID dishes \citep{Bacon:2018dui}. The MICE catalogue, which extends to redshifts of $z=1.4$ will reach even larger beam sizes of $\theta_\text{beam} = 2.36^\text{o}$. Figure \ref{NoiseCl} shows how the beam effect can present challenges in that it causes instrumental noise to dominate at small scales and potentially destroys information there. We will include the beam scale in terms of multipole $\ell_\text{beam}$ on all our power spectra plots (as done in Figure \ref{NoiseCl}) as this is one of the most dominant effects on our results and on \hi intensity mapping power spectra in general.

\begin{figure*}
\subfloat
	{\includegraphics[width=2\columnwidth]{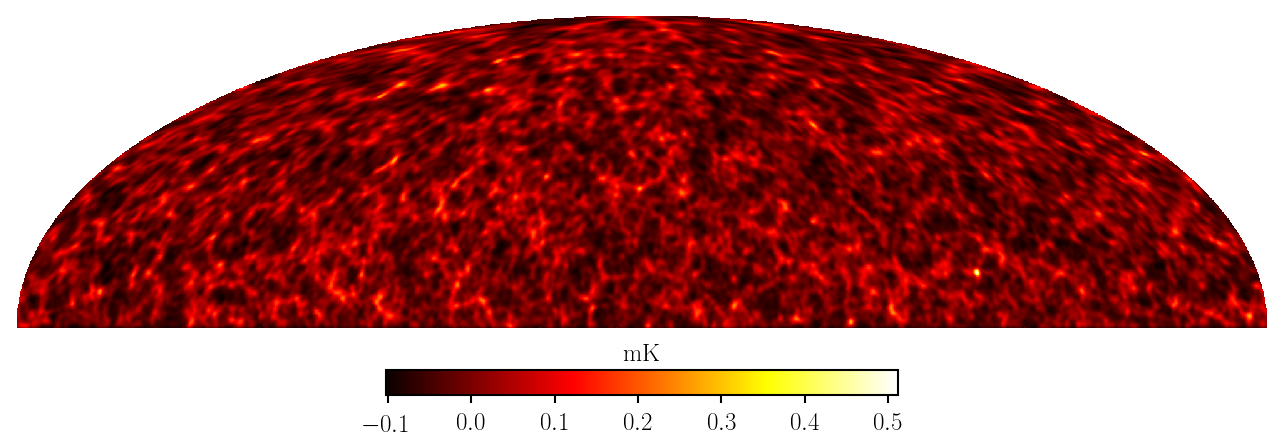}}
    \caption{$\delta T_\hiindex$ intensity map at redshift  $z=0.25$ ($\nu = 1136$MHz) binned using constant redshift intervals of $\Delta z=0.02$. This includes the effects of SKA-like noise and beam, outlined in Sections \ref{noiseintro} and \ref{beamintro} respectively. At this frequency the beam size is approximately $\theta_\text{beam}=1.23^\text{o}$. This example is done with the GAEA catalogue covering half of the sky ($f_\text{sky}=0.5$). This example does not include any foreground contamination.}
\label{dT_HI-map-GAEA}
\end{figure*}
\begin{figure*}
\subfloat
  	{\includegraphics[width=2\columnwidth]{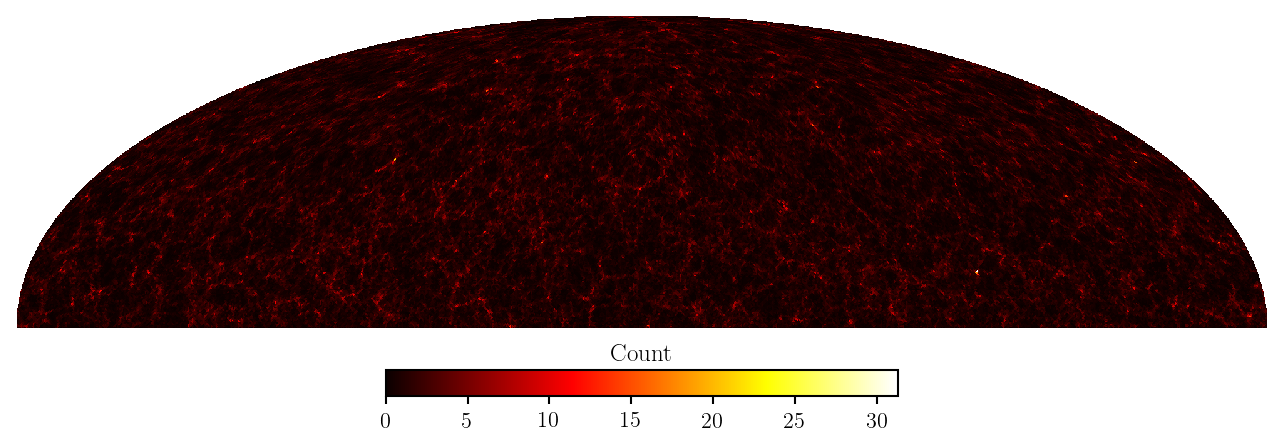}}
    \caption{$n_\text{g}$ optical galaxy number density field with galaxies binned by true redshift at $z=0.25$ with $\Delta z=0.02$. Unlike the intensity map in Figure \ref{dT_HI-map-GAEA}, this map has no beam smoothing since it represents observations by an optical telescope. However, for this demonstration map only, we have downgraded the HEALPix resolution to $\texttt{nside} = 128$. This is to make the shared structure between this and the intensity map at the same redshift more apparent.}
    \label{n_g_z-map-GAEA}
\end{figure*}

An example of a completed intensity map tomographically sliced and collapsed into a 2D angular map is shown in Figure \ref{dT_HI-map-GAEA}. For all our full-sky maps we use HEALPix maps \citep{Gorski:2004by} where the pixelization ensures that each pixel covers the same surface area as every other pixel. We handle the maps in HEALPix RING ordering scheme with resolution $\texttt{nside} = 512$, which corresponds to $12 \times 512^2 = 3,145,728$ pixels across the sky.

\subsection{Optical Galaxy Catalogue Simulation}\label{PhotozcatSec}

For probing large-scale cosmic structure with resolved optical galaxies we use number density fields. 

While we ideally require a simulated catalogue with high number density and completeness for the \hi intensity maps, it would be unrealistic to expect every one of the low mass galaxies to be resolved and detected by a conventional wide area optical survey. Therefore to make this a realistic test we need to introduce some detection threshold which results in only the brightest galaxies being included in our optical sample. We also desire to have realistic $N(z)$ redshift distributions which tail off at higher redshifts where resolved detection becomes more difficult. The way this is all achieved is by invoking a model redshift distribution, given by 
\begin{equation}\label{dNdzModel}
	\frac{dN_\text{g}}{dz} = z^\beta \text{exp}(-(z\alpha/z_\text{m})^\gamma)
\end{equation}
where we use $\alpha = \sqrt{2}$, $\beta=2$ and $\gamma=1.5$ \citep{Harrison:2016stv} which are values typical of stage-IV optical large scale structure survey such as LSST or \euclid. $z_\text{m}$ is the mid-redshift for the particular simulated catalogue we are applying this to e.g. for MICE this would be $z_\text{m}=0.7$. We make the optical samples conform to this distribution by ordering galaxies by stellar mass in each redshift bin. Here we are using stellar mass as a crude approximation of optical brightness which for our purposes will be sufficient. We then pick the `brightest' galaxies in each redshift bin until the model redshift distribution is achieved. This process gives final galaxy catalogues with $2.67 \times 10^7$ galaxies for GAEA, which is an average density of around 54 galaxies per square degree for each of the 24 redshift bins we use. For MICE we achieve a much denser catalogue with $3.97 \times 10^8$ galaxies over a smaller sky area giving $3.2\times 10^3$  galaxies per square degree.

Our optical sample makes no consideration of any classifications of galaxies. All are treated as point-like and either `observed' or not. More investigation could be taken into certain classifications e.g. by colour; red and blue galaxies are expected to cluster differently and have different densities at different redshifts. This could plausibly have an effect on our studies and bias the correlation function; this has been touched upon in a recent cross-correlation study using Parkes \hi intensity maps and 2dF galaxies \citep{ParkesIMxOptDetection}.

Figure \ref{n_g_z-map-GAEA} shows an example of a final over-density field for our optical data. This has been made using the GAEA catalogue at $z=0.25$; similarities between this and Figure \ref{dT_HI-map-GAEA} should be apparent since these are both for the same dataset at the same redshift. We have shown this map with $\texttt{nside} = 128$ to make the clustering pattern more obvious.

\section{21cm Foregrounds \& Their Simulation}\label{FG}

We test the effects on \hi intensity maps of four main foregrounds: \newline

\noindent(i) Galactic synchrotron \newline
\noindent(ii) Extragalactic point sources \newline
\noindent(iii) Galactic free-free emission \newline
\noindent(iv) Extragalactic free-free emission \newline

\noindent Each of these processes emit signals in the frequency region of the redshifted \hi signal i.e. $\sim 1420/(1+z)$ MHz. Each of them are dominant over the \hi signal which is inherently weak. In some cases, such as galactic synchrotron, the foregrounds can be several orders of magnitude higher in observed brightness temperature. 
It is therefore immediately apparent that this a major challenge for the success of the \hi intensity mapping technique.

\begin{figure*}
	{\includegraphics[width=2.1\columnwidth]{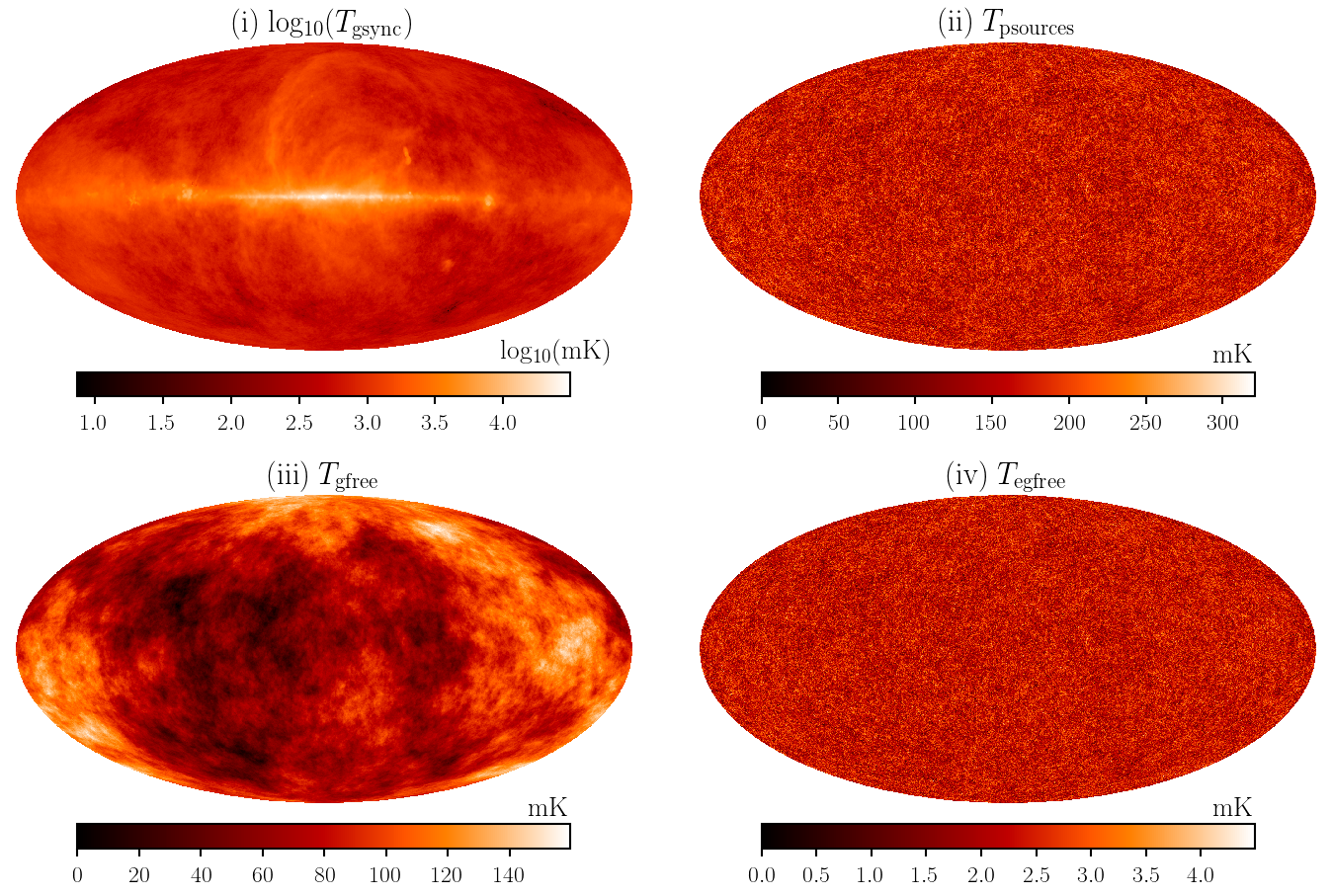}}
    \caption{Full sky maps of each simulated foreground at a frequency of $1136$ MHz ($z=0.25$). These examples do not include noise or beam smoothing. All temperatures are in mK but the galactic synchrotron map (i) shows the logarithm of the temperatures.}
\label{FGmaps}
\end{figure*}

Extragalactic point source foregrounds (ii) are caused by objects beyond our own galaxy emitting signals with wavelengths similar to the redshifted 21cm signal, a typical example being AGNs. (iii) \& (iv) represent free-free emission which is caused by free electrons scattering off ions without being captured and remaining free after the interaction. In this weak-scattering interaction low-energy photons are produced which can enter the $21(1+z)$ cm wavelength window we are interested in. These free-free interaction signals can be detected both within (galactic free-free) and beyond (extragalactic free-free) our own galaxy. 

Lastly the synchrotron emission (i) occurs when high-energy electrons are subject to an acceleration perpendicular to their velocity by the application of a magnetic field. This foreground is typically caused by relativistic cosmic ray electrons accelerated by the galactic magnetic field. It is the galactic synchrotron which is by far the most dominant of the foreground types and is therefore the one we would like to concentrate most on removing.

\subsection{Galactic Synchrotron}\label{galsyncsec}

While it would be fairly straightforward to simulate Gaussian realizations of galactic synchrotron from a model power spectrum, it is far more robust to make use of existing data and use this to emulate the shape of the emission on the sky. This also allows us to study the impact of subtracting a foreground which has wide structures, potentially eliminating information at large angular scales.

Unfortunately, foregrounds within the frequency range of the redshifted 21cm signal ($400 \, \text{MHz} < \nu < 1420$ \, MHz) are less well studied than  other foregrounds, for example those which impact the microwave background emission at higher frequencies ($\nu>10$GHz). Therefore, obtaining actual data maps of galactic emission at regular frequency intervals in the range we are interested is challenging. 

Following a method which has been used in similar \hi foreground studies \citep{Shaw:2013wza, Wolz:2013wna, Alonso:2014sna} we use the Global Sky Model (GSM) \citep{Zheng:2016lul} to generate maps $T_{1420}(\vec{\theta})$ and $T_{400}(\vec{\theta})$ which are emission maps at $1420$ MHz and $400 $ MHz, then use these to construct a full-sky spectral index given by
\begin{equation}
	\alpha(\vec{\theta}) = \frac{\text{ln}T_{1420}(\vec{\theta}) - \text{ln}T_{400}(\vec{\theta})}{\text{ln}1420 - \text{ln}400} \, .
\end{equation}
This is then used to extrapolate the Haslam map \citep{Haslam1982}, which is one of few all-sky maps for galaxy emission around these frequencies,
\begin{equation}
	T_0(\vec{\theta}, \nu) = T_\text{Haslam}(\vec{\theta})\left(\frac{\nu}{408 \, \text{MHz}}\right)^{\alpha(\vec{\theta})}.
\end{equation}
This can now be used to simulate a map of the sky at any desired frequency. However, since the Haslam map does not provide information beyond its own resolution ($\sim 1^\text{o}$), we need a further process to improve the resolution of these maps for any meaningful investigation of small scales. 

We add in this additional small scale information through Gaussian realizations of an angular power spectrum which models galactic synchrotron emission. Following \citet{Santos:2004ju} we make this construction using the angular power spectrum
\begin{equation}\label{FGpowerspec}
	C_\ell(\nu_1,\nu_2) = A \bigg(\frac{\ell_\text{ref}}{\ell}\bigg)^\beta\bigg(\frac{\nu^2_\text{ref}}{\nu_1\nu_2}\bigg)^\alpha\text{exp}\bigg(-\frac{\text{log}^2(\nu_1/\nu_2)}{2\xi^2}\bigg)	\, ,
\end{equation}
where $\xi$ is a parameter which regulates the spectral smoothness of the foreground such that smaller $\xi$ cases are less smooth in frequency and are therefore more of a challenge to disentangle from the cosmological signal. The rest of the parameters are defined in Table \ref{FGparams}. Figure \ref{FGmaps}(i) shows a full-sky map of the simulated galactic synchrotron emission for a frequency slice. 

\begin{table}
	\centering
	\begin{tabular}{lcccr} 
		\hline
		Foreground & A & $\beta$ & $\alpha$ & $\xi$ \\
        \hline
		Galactic synchrotron & 700 & 2.4 & 2.80 & 4.0\\
		Point sources & 57 & 1.1 & 2.07 & 1.0\\
		Galactic free-free & 0.088 & 3.0 & 2.15 & 35\\
		Extra-galactic free-free & 0.014 & 1.0 & 2.10 & 35\\       
		\hline
	\end{tabular}
    \caption{Parameter values for foreground $C_\ell$'s (see Equation \eqref{FGpowerspec}) with amplitude $A$ given in mK$^2$. Pivot values used are $\ell_\text{ref} = 1000$ and $\nu_\text{ref} = 130 \, \text{MHz}$ as per \citet{Santos:2004ju}.}
    \label{FGparams}
\end{table}

Galactic synchrotron has the added complication of being partially linearly polarized. This polarized portion can undergo Faraday rotation which changes the polarization angle of the radiation.  The consequences for the \hi signal have been studied in \citet{Jelic:2010vg, Jelic:2008jg,Moore:2013ip}. Generally speaking this polarization response tends to erode the spectral smoothness of the signal, since it is a frequency dependent effect, and the induced spectral structure is problematic for separating the foreground from the cosmological \hi signal. This requires excellent instrumental calibration to avoid leakages of the polarization effects. The simulation of such polarization leakage is complex and instrument specific. For this work we do not simulate any polarization of the synchrotron emission, but we do opt to convolve all our maps at differing frequencies to a common resolution based on the maximum size of the instrument beam. This is thought to be an active step in mitigating the effects of polarization leakage and is something that is carried out in the Green Bank Telescope \hi intensity mapping data analysis in \citet{GBTHIdetection2}.

\subsection{Point Sources \& Free-Free Emission}\label{psourcefreeSec}

While galactic synchrotron dominates over all other \hi foregrounds, it is still important to consider these additional contaminants since they still dominate over the \hi signal. Extragalactic point sources and extragalactic free-free emission are isotropic in nature, since they are sources beyond our own galaxy. Therefore it is realistic to simulate them with full-sky Gaussian realizations of the angular power spectrum we laid out in Equation \eqref{FGpowerspec} using parameters from Table \ref{FGparams}. This makes the assumption that the source of these foregrounds is Gaussian and also that there is no angular correlation between point sources and \hi emitting galaxies. While point sources will cluster with the underlying matter density, the continuum signals they emit mean that in any one redshift bin, angular correlation between point source signal and \hi is likely to be small.

Galactic free-free emission is not expected to be perfectly isotropic and will have some galactic latitude dependence. However, because it has a low amplitude and very smooth frequency dependence, this will not be a difficult foreground to subtract and we therefore do not believe a more robust modelling is needed here.

For these three foregrounds, point sources, galactic free-free and extra-galactic free-free, we therefore use Equation \eqref{FGpowerspec} and the parameters from Table \ref{FGparams} for our simulations. Figures \ref{FGmaps}(ii),(iii) and (iv) shows maps of these three different foregrounds using the isotropic Gaussian realization approach we have outlined. The lack of galactic latitude dependence is immediately apparent in contrast to the galactic synchrotron map in Figure \ref{FGmaps}(i).

\begin{figure}
  	\includegraphics[width=\columnwidth]{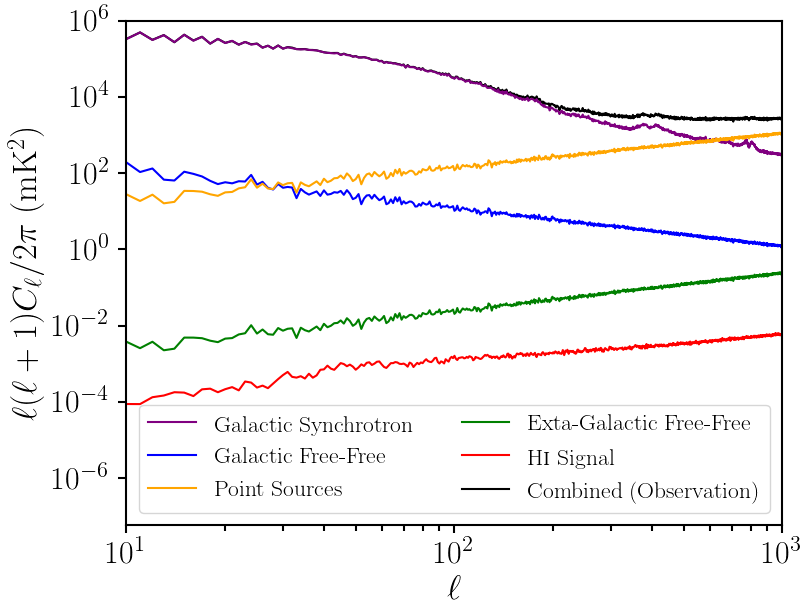}
    \caption{Angular power spectra for all the different simulated foregrounds, and the \hi cosmological signal produced using the GAEA catalogue. The black solid line represents the combined signal from all foregrounds and the \hi cosmological signal. All are at a frequency of 1136MHz ($z = 0.25$) and noise free with no beam effects added.}
    \label{Cl_FG}
\end{figure}

To complete this discussion on \hi foregrounds we include the angular power spectra measured for each of the produced foregrounds in Figure \ref{Cl_FG} along with the cosmological signal. This immediately highlights the challenge faced when attempting foreground subtraction as it demonstrates how dominant all the foregrounds are over the cosmological signal we are trying to extract.

\subsection{Simulated Observable Signal}

To summarize, our simulated sky signal is a composition of maps at certain frequencies (equivalently, redshifts) which can be described by
\begin{equation}\label{dTsky}
    \delta T_\text{sky}(\nu) = \delta T_\hiindex(\nu) + \sum_i \delta T^\text{FG}_i(\nu)
\end{equation}
where the first term comes from the signal described in Section \ref{HIIMSec} and the second term is the contribution from all the different foregrounds outlined previously. Once these maps are combined we smooth the total temperature map $\delta T_\text{sky}$ using the Gaussian beam given by Equation \eqref{beamequation}. We then add the simulated random noise from Equation \eqref{NoiseEq} to emulate basic instrumental systematics, resulting in our final simulated observation $\delta T_{\rm obs}$:
\begin{equation}\label{dTobs}
    \delta T_\text{obs}(\nu) = \textbf{S}_\text{beam}\left(\delta T_\hiindex(\nu) + \sum_i \delta T^\text{FG}_i(\nu)\right) + \delta T_\text{noise}(\nu)
\end{equation}
where $\textbf{S}_\text{beam}$ is the smoothing (or convolution) function.

\section{Foreground Removal}\label{FGremoval}

\begin{figure}
  	\includegraphics[width=\columnwidth]{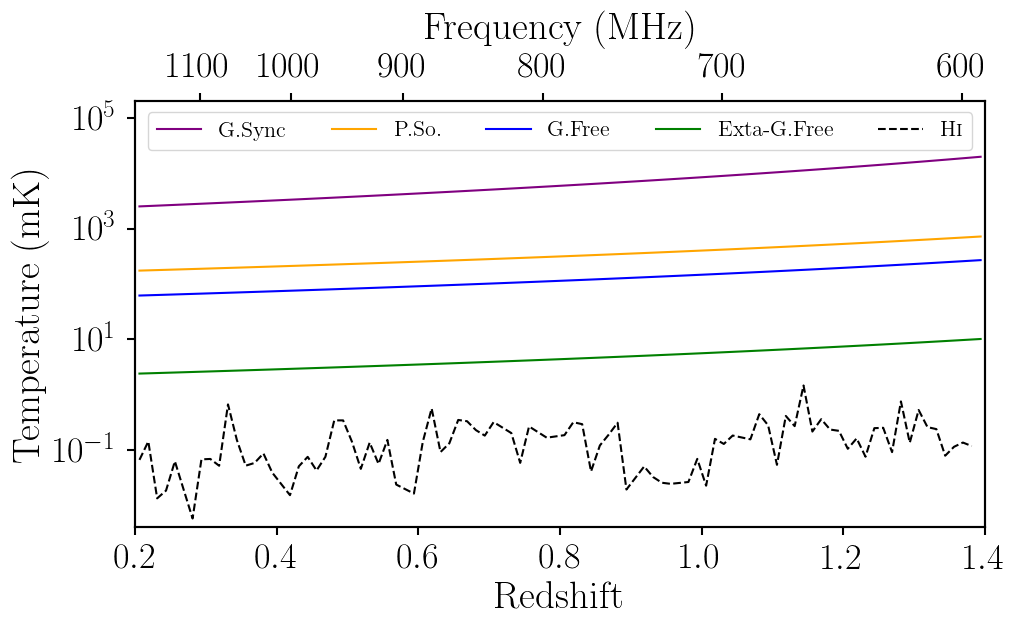}
    \caption{Observed brightness temperatures along a chosen LoS through frequency (or redshift). This is presented for the MICE catalogue with 100 redshift bins to show a large frequency range. The plot demonstrates the foreground smoothness in frequency (coloured solid lines), in contrast to the highly oscillatory fluctuations of the \hi signal (black dashed line).}
    \label{FGLoStemps}
\end{figure}

While foregrounds pose a huge problem for the prospects of exploring cosmology with \hi intensity mapping data, there are some features that help distinguish them from the cosmological 21cm signal. We can utilize the spectral smoothness of the foregrounds to separate them from the \hi, which fluctuates with frequency. Figure \ref{FGLoStemps} shows that along a LoS, the foregrounds are very smooth, whereas the expected signal from \hi has a strong frequency dependence. It is this property that is utilized in a class of methods referred to as blind foreground subtraction. Less general `non-blind' approaches would involve precise modelling of the foreground contamination. Given the lack of data for these foreground signals at the relevant frequencies, this approach is not currently viable.

It is apparent however, that a foreground clean based on this distinguishing spectral smoothness would be more successful for small wavelength radial modes, whereas for larger wavelength radial modes the \hi signal is more similar to the foregrounds. Hence these types of foreground cleans can render large Fourier radial modes (or small $k_\parallel$) useless. Removing large scale modes from \hi intensity maps is therefore an expected effect of a foreground clean and was used as a toy model to emulate the effects of foreground cleaning in \cite{Cunnington:2018zxg}. In the present work we extend our foreground investigation by directly contaminating the maps with the foregrounds we outlined in Section \ref{FG}, and then use state-of-the-art foreground removal techniques to recover our \hi input data and study the impact this will have on fundamental cosmological measurements.

There are several blind foreground removal techniques, for example principle component analysis (PCA) and independent component analysis (ICA) whose distinctions are outlined in \citet{Alonso:2014dhk}. Further blind component separation methods include Generalalized Morphological Component Analysis (GMCA) \citep{Chapman:2012pn} and Generalized Internal Linear Combination (GnILC) \citep{Remazeilles:2011ze}. For this work we examine the \fastica method \citep{Hyvrinen1999FastAR, Chapman:2012yj, Wolz:2013wna, Wolz:2015lwa}, which we describe in the following section. \citet{Alonso:2014dhk} found there to be very little distinction between a PCA and ICA approach to foreground cleaning, so our choice of \fastica as a foreground removal process should not affect the generality of our conclusions.

\subsection{FASTICA Formalism}\label{FasticaFormalismSec}

Here we introduce the basic principles of the Fast Independent Component Analysis (\size{7.5}{FASTICA}) technique, which we will utilize for foreground removal. For a more complete derivation and discussion we refer the reader to \citep{Hyvrinen1999FastAR}. In a blind foreground removal problem we assume that a raw observed signal, such as that outlined in Equation (\ref{dTobs}), can be generalized into a linear equation where the elements making up the signal are statistically independent. That is
\begin{equation}\label{ICAequation1}
	\textbf{x} = \textbf{A}\textbf{s}\, .
\end{equation}
The dimensions and basic description of each term in this equation are given as: \newline
\\
\noindent \makebox[1.5cm]{$\textbf{x}$ $[N_z,\ 1]$}: combined observed signal\par
\noindent \makebox[1.5cm]{$\textbf{A}$ $[N_z,\ m]$}:  mixing matrix - determines the amplitudes of $\textbf{s}$ \par
\noindent \makebox[1.5cm]{$\textbf{s}$ $[m,\ 1]$}:  independent components (containing foregrounds)\newline
\\
Practically this system will have some trace residuals which have some frequency dependence which will include instrumental noise, any residual foregrounds which cannot be classified into an independent component (IC), and the cosmological \hi signal.
\fastica aims to solve Equation (\ref{ICAequation1}) and identify each IC so that from the remaining residual, the \hi signal can be reconstructed. 
For each LoS, sorted into $N_z$ redshift bins and assuming $m$ independent components (ICs) are present, \fastica assumes
\begin{equation}\label{ICAequation2}
	\textbf{x} = \textbf{A}\textbf{s} + \epsilon = \sum_{i=1} ^{N_\text{IC}=m} \textbf{a}_i s_i + \epsilon \, ,
\end{equation}
with $\epsilon[N_z,\ 1]$  the residual (containing \hi signal and noise).
\newline
\\
Under the assumption that each independent component $s_i$ is statistically independent, \fastica attempts to solve Equation (\ref{ICAequation2}) by utilizing the central limit theorem. This states that the greater the number of independent variables in a distribution, the more Gaussian that distribution will be i.e. the probability density function (PDF) of several independent variables is always more Gaussian than that of a single variable. Hence, if we can maximize any statistical quantity that measures non-Gaussianity, then we can identify statistical independence and form a prediction for $\textbf{a}_i$ and $s_i$.

The parameter $m$ must be pre-specified before calculations. This is the number of ICs that can be described by unique non-Gaussian descriptions and is not necessarily the number of different foregrounds one is aiming to find. It is typically assumed that $m\approx 4$ and \fastica then works by obtaining 4 data vectors which are as statistically independent as possible. With {\size{7.5}{FASTICA}}, going to a higher number of ICs than is required converges to the same result. However, the computational cost is increased so for efficiency, the lowest value for $m$ which gives the best possible result is sought.

The \fastica process considers all LoS simultaneously. Therefore for its calculations on maps with a number of pixels given by $N_\text{pix}$, the ICs $\textbf{s}$ in Equation \eqref{ICAequation2} are actually maps, and hence an array with size $[m,\ N_\text{pix}]$, while $\textbf{x}$ and $\epsilon$ are arrays of size $[N_z,\ N_\text{pix}]$. Furthermore, as we will further explain below, \fastica involves some expectation value calculations which rely on a number of samples and for this it uses the $N_\text{pix}$ different LoS.

\begin{figure*}
\subfloat
	{\includegraphics[width=2.1\columnwidth]{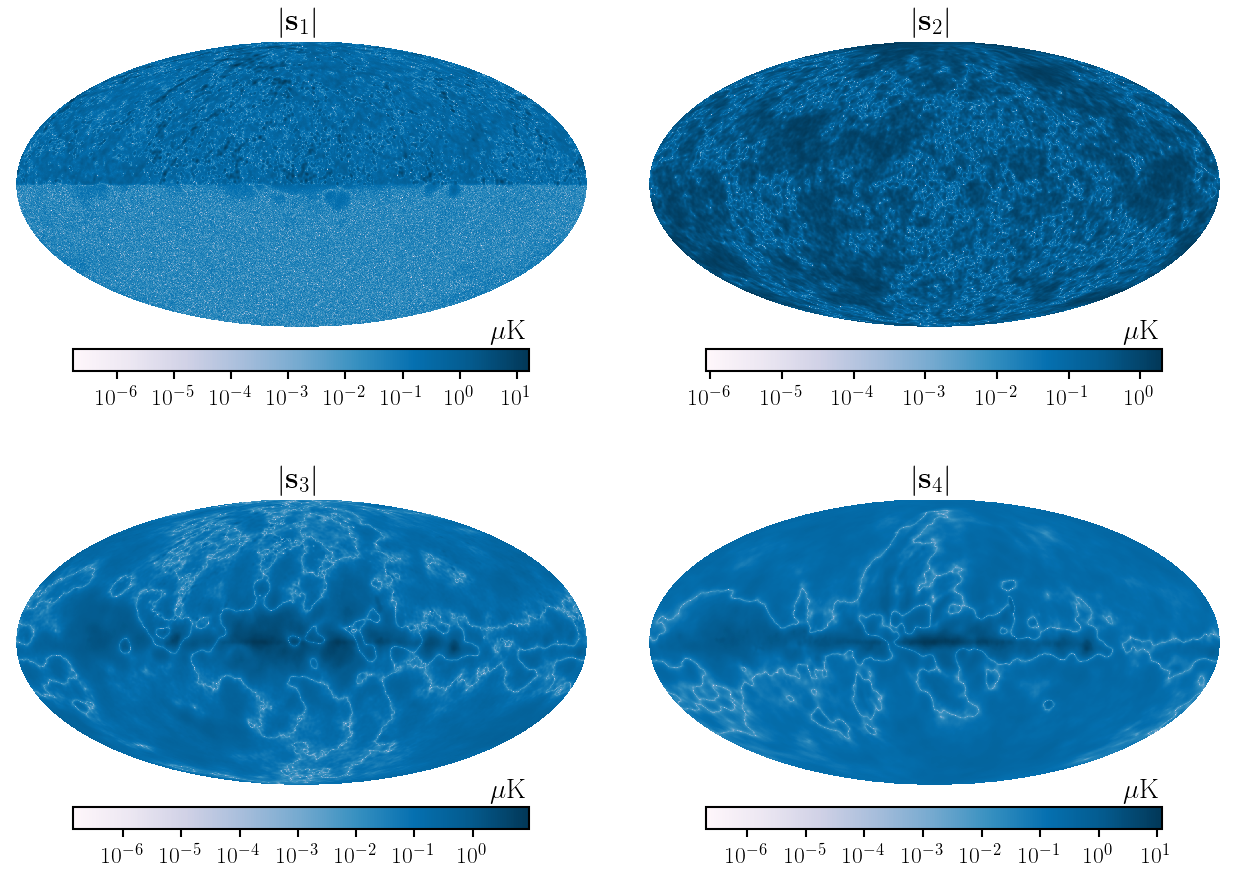}}
    \caption{Independent component maps found using \fastica with $m=4$ on the GAEA simulation contaminated with foregrounds. This is for a constant beam of $\theta_\text{beam}=0.5^\text{o}$ at all frequencies. Temperature fluctuations are given in $\mu$K but the true amplitudes for the estimated foregrounds are determined by their combination with the mixing matrix.}
\label{ICmaps}
\end{figure*}

\begin{figure}
    \centering
  	\includegraphics[width=0.9\columnwidth]{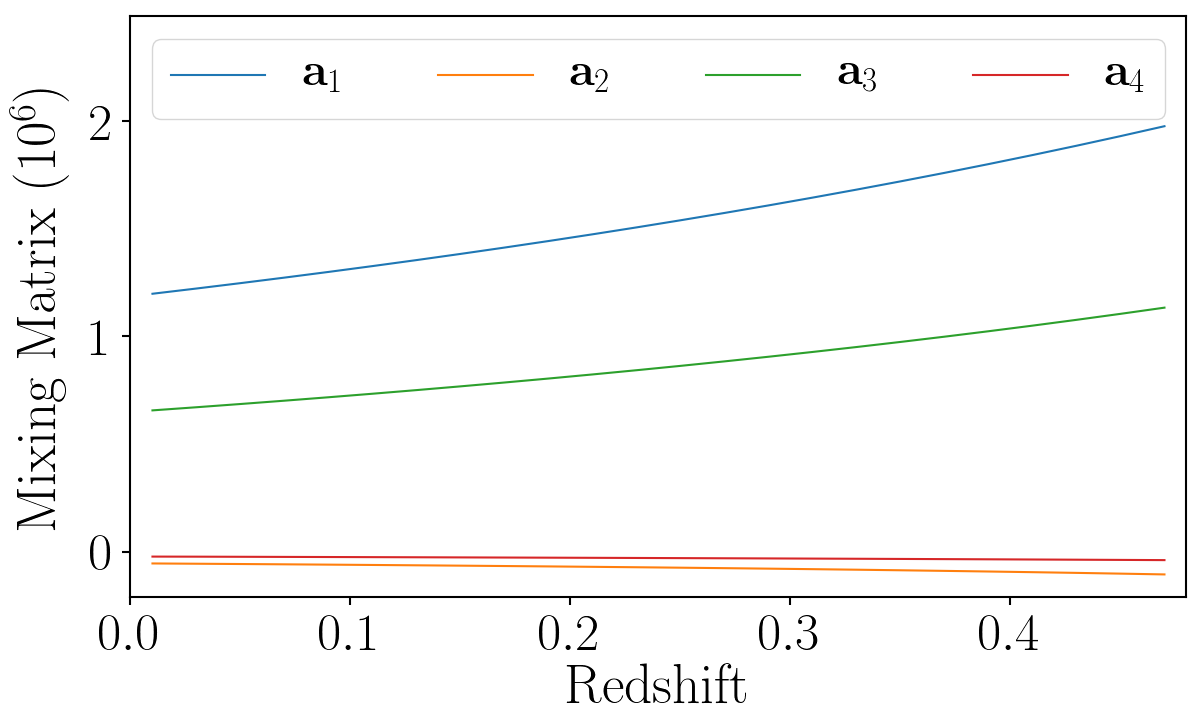}
    \caption{Mixing matrix elements as outlined by Equation (\ref{ICAequation2}). Combination of these with the independent components in Figure \ref{ICmaps} determines the subtraction to be made from the combined observed signal at each frequency.}
\label{MixingMatrix}
\end{figure}

To obtain \textbf{s} we start by inverting Equation (\ref{ICAequation2}), ignoring the residual term $\epsilon$ which will just be left over from signal not contained within the ICs. We can therefore write
\begin{equation}\label{ICAinverse}
	\textbf{s} = \textbf{W}\textbf{x}\, ,
\end{equation}
here $\textbf{W}$ is the weighting matrix, defined as the inverse of $\textbf{A}$ in Equation~(\ref{ICAequation1}). Under the assumption that the elements $\textbf{s}$ are as statistically independent as possible, \fastica then begins maximizing the non-Gaussianity. For a measure of Gaussianity it uses negentropy $J(y)$, which for a variable $y$, is based on typical entropy $H(y)$ defined as 
\begin{equation}
	H(y) = -\sum_i P(y=a_i)\log P(y=a_i) \, ,
\end{equation}
where $P(y=a_i)$ is the probability that $y$ equals a possible value  $a_i$. The modification made to obtain the negentropy $J(y)$ is
\begin{equation}
	J(y) = H(y_\text{G}) - H(y) \, ,
\end{equation}
where $y_\text{G}$ is a unit-variance Gaussian random variable. In practice, negentropy is computationally hard to calculate and requires numerous realizations to obtain information on  probability distributions. However, using the maximum entropy principle, we can write 
\begin{equation}\label{negentropyequation}
	J(y) \approx -\sum_i^n k_i [\langle G_i(y) \rangle_\theta - \langle G_i(y_\text{G}) \rangle_\theta] \, ,
\end{equation}
where $k_i$ are positive constants, $G_i$ is referred to as the contrast function, and all pixels are utilized by averaging over them (this is denoted by $\langle \rangle_\theta$). For the contrast function, whilst practically any non-quadratic function will work, \fastica mainly uses
\begin{equation}
    G_1(y) = \frac{1}{a_1}\log \cosh(a_1 y),\ G_2(y) = -\frac{1}{a_2}\exp(-a_2 y^2 /2) \, ,
\end{equation}
where $1\leq a_1 \leq 2$ and $a_2\approx 1$.\newline
\\
\noindent In a nutshell, \fastica delivers a method of reconstructing the foreground signals as $m$ ICs and then the residual $\epsilon$ between this reconstruction and the raw observed input map is the recovered cosmological \hi signal plus any receiver noise and residual foreground contaminants. A final point to include is that the mean temperature of the \hi cosmological signal is a smooth function of frequency and is therefore incorporated into the ICs of the analysis. This information is therefore lost and the residual maps are required to be renormalised to some model mean temperature or treated as $\delta T$ observables as in Equation (\ref{deltaTeq}). 
\subsection{FASTICA Results}

Here we seek to validate the \fastica reconstruction process introduced in the previous Section \ref{FasticaFormalismSec} by presenting results from our simulations outlined in Section \ref{FG}. Since neither of our cosmological simulations cover the full sky, we only add and remove foregrounds to the footprint covered by GAEA and MICE. Restricting the foreground analysis to these patches represents a more realistic emulation of a cosmological survey. However, we found no noticeable difference when we conduct the foreground removal over the full sky compared with conducting it over the cosmological simulation footprint.

Figure \ref{ICmaps} shows the IC maps found after \fastica has been applied. This is the only occasion where the foreground analysis is done for the full sky and we have chosen to do this purely for demonstrative purposes of the \fastica process. It is interesting to note that the third and fourth ICs clearly seem to pick up the galactic synchrotron angular shape whereas the second IC shows structure across the sky. The first IC is largely contained in the top half of the map, where the \hi cosmological signal lies for the GAEA catalogue. This suggests that it is this component which is collecting large radial modes which belong to the cosmological signal along with the $\overline{T}_\hiindex$ average which smoothly fluctuates and therefore is removed. Despite trying a number of different values of $m$ (the number of ICs) it appears that it is always the case that some cosmological signal will be removed. These ICs from Figure \ref{ICmaps} are then combined with the mixing matrix (displayed in Figure \ref{MixingMatrix}) as described in equation \eqref{ICAequation2}.

\begin{figure*}
  	\includegraphics[width=\textwidth]{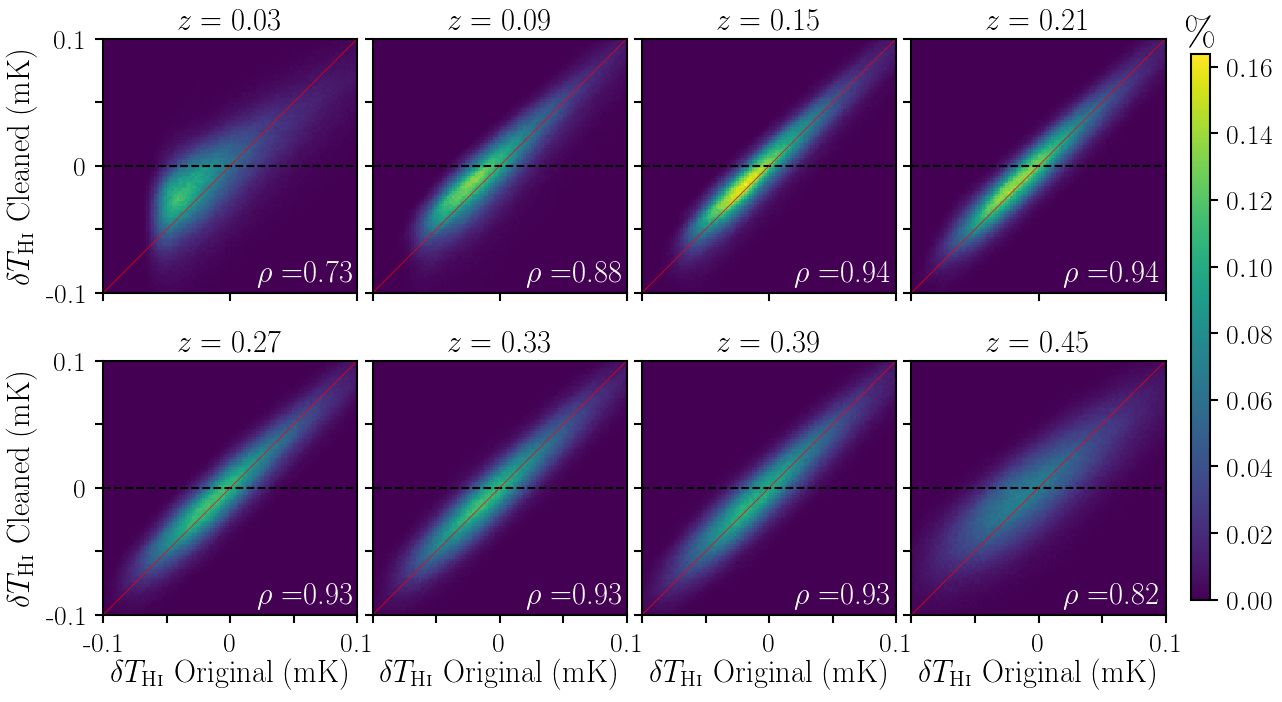}
    \caption{Histogram showing the original \hi temperature against the \fastica reconstructed value for each pixel in a range of redshift bins for the GAEA model. Each histogram has been normalized such that the histogram values sum to 100\%. We also include the Pearson correlation coefficient $\rho$ for each redshift to quantify the agreement. For a perfectly working foreground clean we would expect an entirely one-to-one ($\rho=1$) agreement along the thin diagonal red line. We can see how \fastica is less effective at extreme ends of redshift range with a wider dispersion of values.}
\label{CleanVOrigHist}
\end{figure*}

Figure \ref{CleanVOrigHist} shows a pixel-by-pixel comparison between original values in the $\delta T_\hiindex$ intensity maps and the cleaned values for some selected redshift bins in our GAEA simulation. For a perfectly performing reconstruction we would obtain all values along the red diagonal line, i.e. all values would match their originals. We can see that this is not the case but largely \fastica is performing reasonably well with a Pearson correlation coefficient of $\rho \geq 0.93 $ for most redshifts. We expect a value of $\rho=1$ for a perfectly performing foreground clean indicating perfect correlation between original and cleaned maps. Figure \ref{CleanVOrigHist} also shows that this method of foreground cleaning performs better at the mid-ranges of redshift. This is not a redshift specific effect since we also see similar results in the MICE model where the best agreement is at redshift $z\sim 0.8$ which is the mid-redshift for its range. This suggests that there are some edge effects in the foreground removal process causing it be less effective at the extreme radial ends of the input data, a result previously noted e.g. \citet{Wolz:2013wna}.

\begin{figure*}
\includegraphics[width=\textwidth]{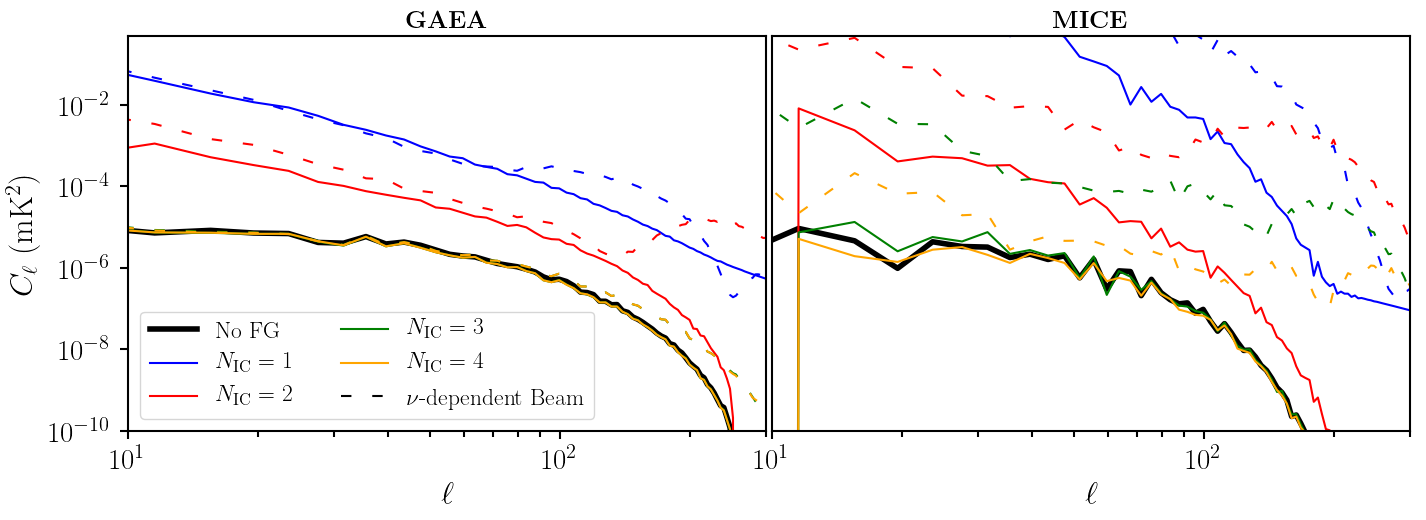}
\caption{Impact of foregrounds on the \hi auto-power spectrum for both the GAEA and MICE catalogues. Thick solid black line shows the original \hi signal with no foregrounds. The coloured lines then show different values of $m$ used i.e. the number of independent components assumed in the \fastica process. We also include the results from using a beam which varies with frequency (dashed lines) and how this damages performance. For GAEA it appears that there is little need to go beyond 3 independent components for a successful reconstruction in the constant beam size case. Whereas for MICE even going to 4 independent components there is still some disagreement at large scales (small-$\ell$). The disagreement we see is caused by residual foreground which \fastica has failed to remove. Similar results have been previously shown in e.g. \citet{Wolz:2013wna}. These results are for mid-range redshifts for each catalogue with $z=0.25$ for GAEA and $z=0.825$ for MICE.}
\label{Clauto_NIC}
\end{figure*}

Figure \ref{Clauto_NIC} indicates how well the \hi auto power spectrum can be recovered with \fastica and shows how varying the number of ICs affects the recovery. We show results from both simulations, and it is interesting to note the difference between the two. We see that with GAEA only 3 ICs are needed for a successful reconstruction, however for MICE even 4 ICs is not sufficient for good agreement at large scales (small-$\ell$). We tested a larger number of ICs with little improvement. The difference in results is probably due to the fact that MICE has a smaller sky coverage (25\% of GAEA) which means less samples to average over for negentropy estimation in equation \eqref{negentropyequation}. Furthermore, since MICE has a deeper redshift range (extending to $z=1.4$ whereas GAEA is only up to $z=0.5$) the constant beam size that we convolve with is much larger for MICE, $\theta_\text{beam}=2.36^\text{o}$ against GAEA's  $\theta_\text{beam}=1.46^\text{o}$. This difference in beam size is also evident from the scales at which the power spectrum seems to degrade. Due to its larger beam, the MICE power spectrum begins to tail off at lower-$\ell$ than GAEA. Lastly this plot also includes results where each tomographic slice has been smoothed by a varying amount due to the frequency dependence of the beam. This is shown as the dashed lines, and it is evident that results are much worse when compared with the constant beam case. This is discussed further in the following section.

\subsubsection{Frequency Dependent Beam Size}

As previously outlined in Section \ref{beamintro}, the intensity maps at different frequencies will have different beam sizes defined by equation \eqref{beamequation}, meaning intensity maps at lower redshift have less degradation of angular scales. However, since \fastica finds $m$ IC maps and then subtracts these from the total observation based on the mixing matrix $A$, trying to obtain e.g. 4 IC maps based on $N_z$ intensity maps with different resolutions for each will cause problems because the IC map resolution will not match each of the intensity maps. This is exactly why we see poorer performance in Figure \ref{Clauto_NIC} in the case where there is a frequency dependent beam size (dashed lines) especially at smaller scales (large-$\ell$) where the beam has a more dominant effect.

The way we resolve this issue is by carrying out a further convolution on the intensity maps such that each tomographic slice is smoothed to the same resolution. We therefore take the maximum beam for the particular redshift range and smooth over all maps with this constant beam size. \fastica then finds IC maps which, when subtracted from the observed signal, prove more effective for reconstructing the original \hi signal as shown by the solid lines in Figure \ref{Clauto_NIC}.

Artificially re-smoothing over all our intensity maps may appear to be a wasteful process in terms of loss of large-$\ell$ modes, but it is necessary for a successful \fastica reconstruction. 
In fact, choosing a common resolution significantly larger than the max beam has additional benefits when dealing with real data, as an effective way of mitigating the effects of polarization leakage. The polarization leakage introduces contaminations on scales of the order of the primary beam, which would negatively affect foreground removal. The convolution of the maps with a beam larger than these scales, smooths over the extra structure and mitigates these effects \citet{GBTHIdetection2}.

\subsubsection{Increasing the Number of Frequency Bins}

For both our GAEA and MICE simulations we are only using 24 redshift (frequency) bins with the bin width determined by a constant separation in redshift $\Delta z$. This may be seen as quite a low number of bins to be using in an intensity mapping simulation which uses an ICA process. This is largely out of necessity due to the choice of simulation approach: since we are using $N$-body simulations we have a finite number of galaxies to use from which to build intensity maps. By using bins which are too thin we risk under-sampling the intensity maps and making them an unrealistic emulation of a continuous field of emission.

In practice when using real data, the typical approach would be to perform the \fastica method on more maps ($>100$ frequency channels), then re-stack these into fewer bins for cosmological analysis and cross-correlations with optical data. We trialed this with our MICE catalogue using 240 bins, and found that it made no improvement on the \fastica foreground removal, hence justifying our choice of using $24$ frequency bins in all our analysis. Furthermore, in cross-correlation with optical imaging surveys, the intensity maps would likely need stacking into thicker bins to match the thick bin widths needed for the poorly constrained photometric redshifts.

\section{HI $\times$ Optical Cosmology with Foregrounds}\label{HIxOpticalSec}

In this section we investigate the impact that \hi foreground contamination and removal with \fastica have on the cross-correlation power spectra $C_\ell^{\text{g},\hiindex}$ with our simulated optical catalogues.

\begin{figure}
  	\includegraphics[width=\columnwidth]{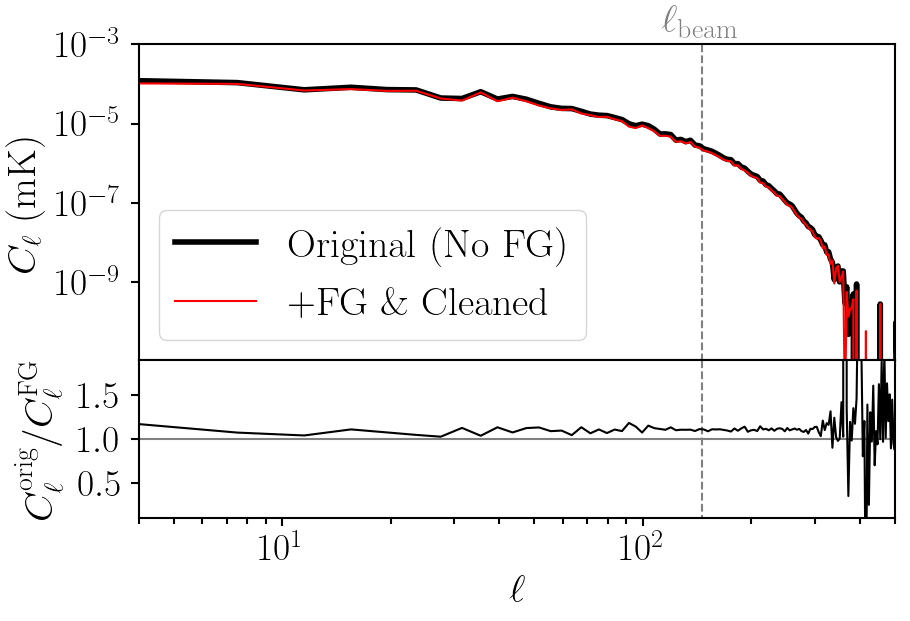}
    \caption{Cross-correlation angular power spectrum between the \hi intensity map at redshift $z=0.25$ and the optical galaxies binned using their true redshifts. This is representative of a scenario in which spectroscopic redshifts are used in the optical survey. The original result with no foregrounds is shown as the black thick line and the case where foregrounds have been included then removed by \fastica is shown as the red thin line. The bottom panel shows the ratio of the two spectra. This test was carried out on the GAEA simulation where the \hi intensity maps have been re-smoothed with a constant maximum beam of  $\theta_\text{beam}=1.46^\text{o}$.}
\label{Cl_cross-zbinned}
\end{figure}

In recent work, \citet{Blake:2019ddd} has developed a framework which models observational effects on 3D power spectra for \hiindex-optical cross-correlations. This framework can be extended to include the effects of foreground removal and photometric redshift uncertainty. By doing this one could analytically model the foreground removal effects, as well as the photometric redshift effects, as a loss of small and large $k_\parallel$ modes respectively, and attempt quantitative corrections accordingly. In this work we aim to use our simulations to investigate what corrections can be made to the data to extract the most information from these cosmological measurements.

To begin exploring how \hi foregrounds can impact cross-correlations with optical surveys we first perform a best-case scenario test and cross-correlate with an optical survey which we assume has very well constrained redshifts; Figure \ref{Cl_cross-zbinned} shows the result of this cross-correlation. Here we bin the optical galaxies from the GAEA simulation by their true redshift with constant bin width of $\Delta z = 0.02$. This is exactly matched to the frequency bins used for the 21cm intensity maps using $\nu = \nu_{21}/(1+z)$), so we have a sample of optical galaxies at $z=0.25$ to cross-correlate with an \hi intensity map at the same redshift. This shows that foregrounds should have little impact on optical spectroscopic cross-correlations. The bottom panel shows a small bias which in principle could be corrected for by constructing a foreground cleaning transfer function \citep{Switzer:2015ria}, but it is encouraging that our initial efforts have already reconstructed the cross-power to within 8.5\% at scales below those unaffected by the beam ($\ell<\ell_\text{beam}$). It is only at higher $\ell$, way below the resolution of the beam ($\ell_\text{beam})$, that we start to have large errors on $C_\ell$. This is unsurprising since this is going beyond the scales of the radio instrument's resolution. We experimented with smoothing the optical field to replicate the \hi intensity map resolution but find no mitigation of the noise we see at $\ell>250$.

\subsection{Optical Redshift Uncertainty}\label{PhotozUncertainty}

Future optical galaxy redshift surveys such as LSST and \euclid will rely on using photometric redshifts for estimating the radial position of each galaxy (note that \euclid will also perform a wide spectroscopic survey). It is therefore important to forecast the cross-correlation potential between  \hi intensity maps and photometric galaxy redshift surveys, taking into account foreground removal effects. The higher uncertainty on redshift measurement inherent in these photometric surveys, equates to a degradation in radial mode measurement on small scales. Since foreground removal also impacts radial modes but on larger scales, it is unclear whether combining these two effects will leave enough useful modes for a cross-correlation signal \citep{Witzemann:2018cdx}.

\begin{figure}
  	\includegraphics[width=\columnwidth]{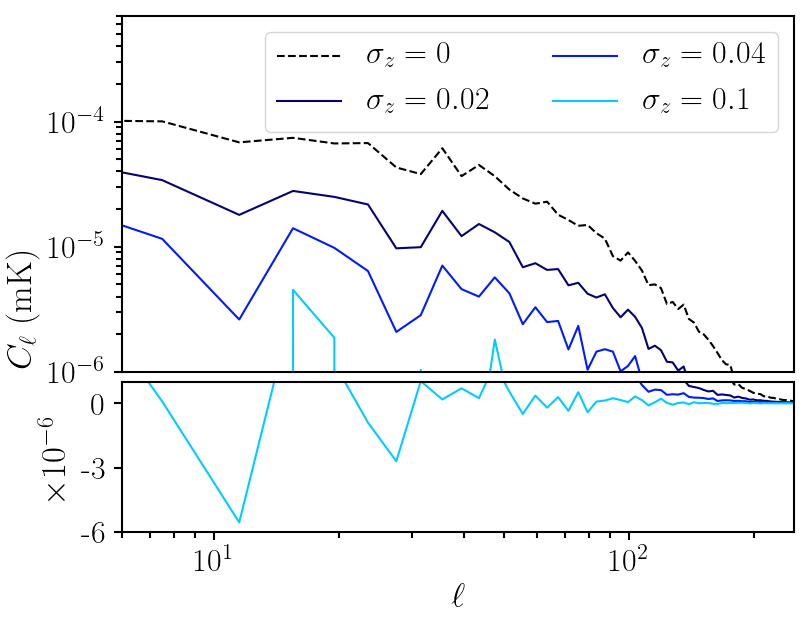}
    \caption{Cross-correlation between \hi intensity maps with \fastica reconstruction and an optical survey using GAEA. We degrade the constraints on the optical galaxy redshifts by increasing the redshift error $\sigma_z$ shown by going from dark to ligher blue. In other words we go from cross-correlating intensity maps with a spectroscopic-like ($\sigma_z \sim 0$) survey, to a photometric-like survey where there is significant uncertainty on the optical galaxy redshifts. This strongly affects the measured cross-correlation power spectrum. Plot includes a hybrid log-linear $y$-axis to fully demonstrate the degradation in power.}
\label{PhotozCls}
\end{figure}

To investigate this we can begin by simply introducing a Gaussian error on our optical redshifts for each galaxy and cross-correlate with foreground contaminated intensity maps. Figure \ref{PhotozCls} shows the effect on the cross-power spectrum when we introduce a Gaussian photo-$z$ error $\sigma_z$ into each of the optical galaxies. We can see how increasing the uncertainty in redshift (dark to light blue lines) rapidly degrades the agreement with the original (black-dashed line) where no redshift error is applied. \citet{Abell:2009aa} suggests a fiducial model of $\sigma_z = \sigma_{z_0}(1+z)$ is appropriate for an LSST-like instrument, where $\sigma_{z_0}=0.05$. Therefore, the fact that Figure \ref{PhotozCls} suggests the cross-power spectra signal-to-noise will be damaged for $\sigma_z \sim 0.1$, which would correspond to LSST's photo-$z$ error at $z=1$, is cause for concern.

We can further explore this with the use of some more robust photometric redshift simulations and compare to foreground free cross-correlations. Realistic photometry for a number of optical surveys is included within the MICEv2 simulation,  for example the Dark Energy Survey (DES)\footnote{\href{www.darkenergysurvey.org/}{www.darkenergysurvey.org/}}. We thus make use of the DES-like photometric redshifts available to make a more robust forecast of the cross-correlation between a photometric survey and \hi intensity maps. We refer the reader to the MICE website\footnote{\href{http://maia.ice.cat/mice/}{http://maia.ice.cat/mice/}} for more details on how these DES-like photometric redshifts were simulated.

\begin{figure}
  	\includegraphics[width=\columnwidth]{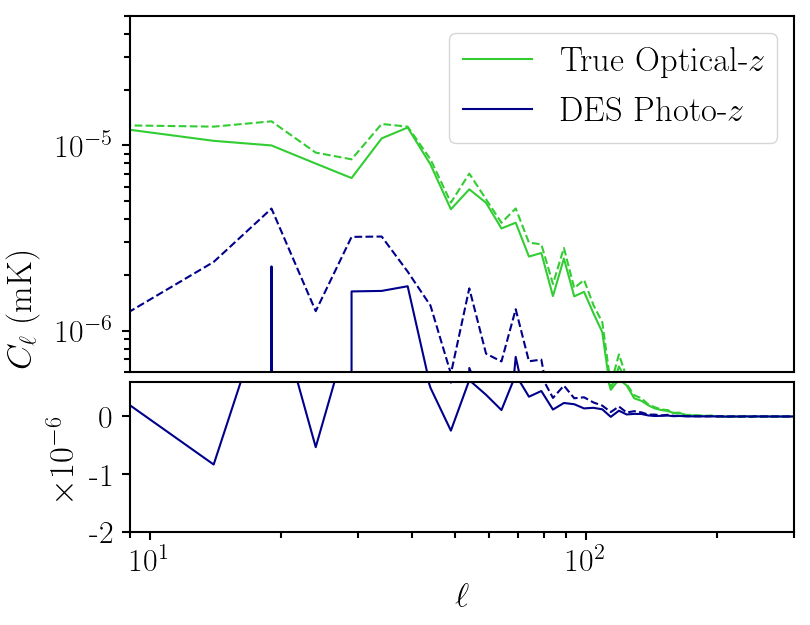}
    \caption{Cross-correlation between \hi intensity maps with MICE optical galaxies. Dashed lines show the cases without \hi foregrounds, solid lines show the impact of including them. We use the DES-like photometric redshifts available in MICE for the photo-$z$ forecasts shown in blue and compare these with using ideal true redshifts (green). While a drop in signal is inevitable when using less constrained redshifts, including the effects of \hi foregrounds (solid lines) degrades the signal further in the photo-$z$ case. These tests have been performed at redshift $z=0.725$.}
    \label{DES-photozCl}
\end{figure}

Figure \ref{DES-photozCl} shows the results when we include these simulated DES-like photometric redshifts in our simulations. The dashed lines show the cross-correlation power spectrum with the original \hi intensity map with no foreground contamination. The solid lines then show the inclusion of foregrounds and a \fastica reconstruction. What is clear from this plot is that while we still get a degradation in signal from using photometric redshifts (blue line) compared with true redshifts (green line), the signal deterioration accelerates in the case where \hi foregrounds are included in the simulation. 

The conclusion from our GAEA simulation using Gaussian photometric redshifts and MICE using DES-like photometric redshifts appears to be the same and both forecast damaging signal loss when \fastica reconstructed intensity maps are cross-correlated with photometric redshift surveys. 

\subsection{Mitigating the Effects of \fastica}\label{CorrectionsSection}

Here we begin investigating the precise reasons why combining the effects of \hi foregrounds and the poor redshift constraints from photometric galaxy surveys is so detrimental to the cross-correlation signal. Generally, it can be considered unsurprising that combining an effect that removes information at large radial modes, with a survey which has poor constraints at small radial modes, can damp the amplitude of projected angular power spectra, as we see in Figures \ref{PhotozCls} and \ref{DES-photozCl}. Our aim here is to quantify this explanation with the hope of being able to provide a solution. 

\begin{figure}
  	\includegraphics[width=\columnwidth]{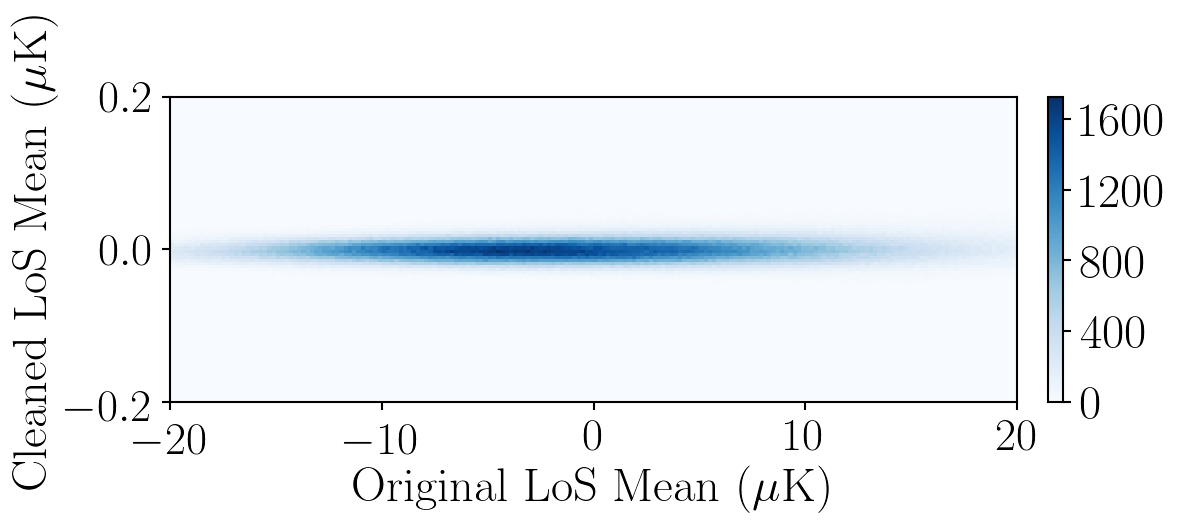}
    \caption{The mean $\delta T$ temperatures along the line-of-sight (LoS) for the original \hi intensity map against one with which has undergone a \fastica foreground clean. This is shown for all available LoS in the GAEA simulation which for $f_\text{sky}=0.5$ and $\texttt{nside} = 512$ equates to over 1.5 million pixels (or LoS). The plot shows how \fastica essentially removes any non-zero LoS mean present in the original \hi signal and collapses it to zero.}
\label{LosMeans}
\end{figure}

It is interesting to look at the effects a foreground clean has along the line-of-sight (LoS) of the \hi intensity mapping data. It is known that large radial modes are removed since this is where the contamination from foregrounds lies due to their smooth variation in frequency. Figure \ref{LosMeans} shows the specific effect this has and illustrates how the foreground clean removes all information on the mean temperature along the LoS. Our simulations are arranged such that the transverse mean of each map is zero but even with this setup it is of course still possible to have a large range of values for the LoS mean temperatures, which is what we see in Figure \ref{LosMeans}. However, we can see that the large range of LoS mean values present in the original \hi signal (shown on the $x$-axis) are removed after the foreground clean to a much narrower range (shown on the $y$-axis). It is worth pointing out that the $y$-axis range is two orders of magnitude smaller than the $x$-axis. So essentially a blind foreground clean will destroy any non-zero mean along the LoS. The original line-of-sight means have a slight skewness away from zero and centre at around $-4 \, \mu$K. This is caused by the presence of some dominant bright pixels which, when setting transverse means in each map to zero, can result in there being more negative temperatures than positive ones.

It is conceivable that an increase in the number of redshift bins could affect this LoS result, so we therefore conducted a test using the MICE catalogue and extended to 240 redshift bins following the same procedure. Even with this more realistic number of redshift bins we still find a similar result to Figure \ref{LosMeans} suggesting that this is not a feature of the relatively low number of redshift bins we are using.

\begin{figure*}
  	\includegraphics[width=\textwidth]{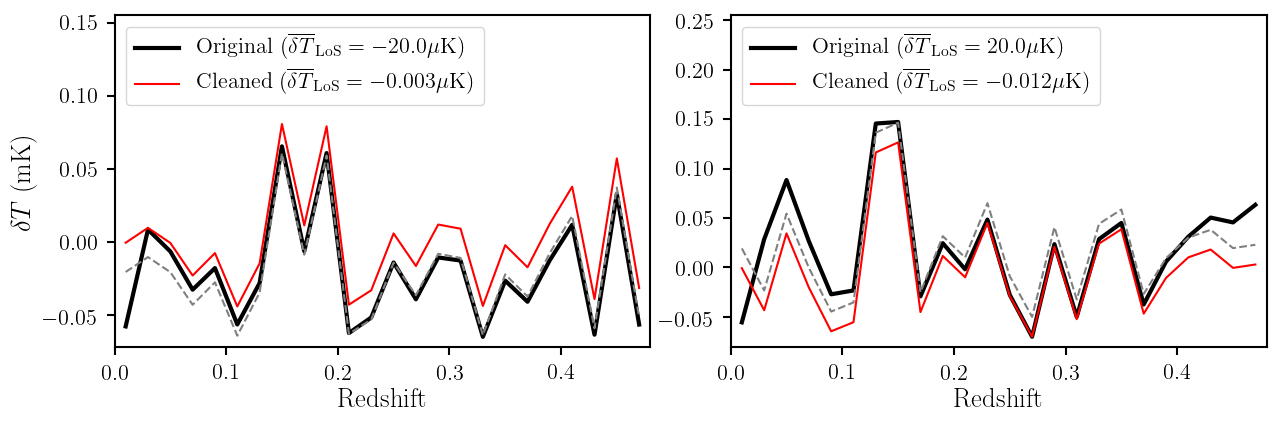}
    \caption{GAEA $\delta T$ amplitudes along chosen lines-of-sight (LoS). Original mean values along the LoS are given in the legend along with the cleaned ones. The thick black line shows the original amplitude and the red solid line shows the impact of a foreground contamination and \fastica foreground clean. The grey dashed line shows the amplitude with the LoS mean added back on as outlined in Equation (\ref{LoSReconstructionEq}).}
\label{LoSReconstruction}
\end{figure*}

\begin{figure}
  	\includegraphics[width=\linewidth]{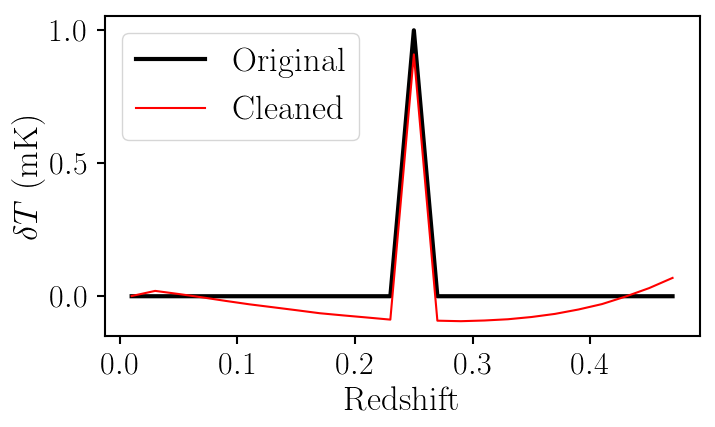}
    \caption{Effect of FASTICA on a test response function. For the GAEA model, all values along a chosen LoS have been set to 0 except one at $z=0.25$ which is set to 1. This data is then subject to a FASTICA clean. An amplitude change from the LoS mean removal is apparent and there are also under-dense side-lobes either side of the temperature spike.}
\label{ResponseFunction}
\end{figure}

In summary, the problem is that while \fastica reconstructs the shape of the LoS signal, unfortunately it changes the amplitudes in an unpredictable manner based on the original LoS mean. The further from zero a particular LoS mean is, the greater the change in amplitude for pixels along this LoS. We attempt to model this by hypothesising that in a blind foreground clean the main resulting change is given by
\begin{equation}\label{LoSLoss}
    \delta T_\text{clean}(\vec{\theta},\nu) \sim \delta T_\text{orig}(\vec{\theta},\nu) - \overline{\delta T}_\text{LoS}(\vec{\theta}) \, ,
\end{equation}
where $\overline{\delta T}_\text{LoS}(\vec{\theta})$ is the mean fluctuation along a LoS for a pixel at position $\vec{\theta}$,
\begin{equation}
    \overline{\delta T}_\text{LoS}(\vec{\theta}) = \frac{\sum_i \delta T_\text{orig}(\vec{\theta},\nu_i)}{N_z} \, ,
\end{equation}
where the summation is over the $N_z$ number of frequency (or redshift) bins. 

Figure \ref{LoSReconstruction} shows the impact along the LoS resulting from the effect outlined in Equation (\ref{LoSLoss}). We have chosen two pixels and show their $\delta T$ values through redshift, taking two extreme examples for demonstrative purposes. The plot on the left is for a pixel where the original LoS mean $\overline{\delta T}_\text{LoS}(\vec{\theta})$ is from the extreme low end from Figure \ref{LosMeans}. The plot on the right is for a pixel with a high $\overline{\delta T}_\text{LoS}$. In both cases their LoS means are collapsed to zero for the reasons discussed above and the impact this has on the agreement between individual values through redshift is evident. 

We can demonstrate that this is the main impact of a blind foreground clean by reversing the effect, i.e. adding back in the original LoS mean to each foreground-removed pixel:
\begin{equation}\label{LoSReconstructionEq}
    \delta T_\hiindex(\vec{\theta},\nu) = \delta T_\text{clean}(\vec{\theta},\nu) + \overline{\delta T}_\text{LoS}(\vec{\theta}) \, .
\end{equation}
The corrected $\delta T_\hiindex$ should agree with the original signal $\delta T_\text{orig}$. We have tested this and find this to be the case and show the results of this approach in Figure \ref{LoSReconstruction}, where we included the reconstructed LoS based on Equation (\ref{LoSReconstructionEq}) shown by the gray dashed line. 

Unfortunately, this LoS \hi mean reconstruction is challenging in reality. The original $\overline{\delta T}_\text{LoS}$ will be information buried in the foreground contaminated maps, and which is then lost after the foreground clean. So performing the process outlined in equation \eqref{LoSReconstructionEq} would require some extra information to reconstruct these LoS means.

In a similar demonstration to Figure \ref{LoSReconstruction}, we also analyse the \fastica result on a test response function in the form of a Dirac-delta spike in temperature, shown in Figure \ref{ResponseFunction}. By manipulating the GAEA data such that all pixels along a chosen LoS are set to 0 except for one which is set to 1, we can gain a deeper insight into the effects of a foreground clean. The large side-lobes which form either side of the temperature spike can explain why the cross-correlation with photometric galaxy data is performing so badly. A galaxy at $z=0.25$ with high measured redshift uncertainty, is likely to cross-correlate with the false under-temperature regions. This effect, compounded over many galaxies and temperature spikes, could cause signal loss.

As an additional problem, we also find that this kind of foreground removal is less successful at the extremes of  the redshift range (something already concluded from Figure \ref{CleanVOrigHist}). Therefore reconstructing the LoS means will not be a sufficient correction on its own at the redshift edges of the data. 

All this highlights the problems for the future success of \hi intensity mapping cross-correlations with photometric galaxies. Nevertheless, photometric galaxy surveys are an important choice of probe to cross-correlate with given their complementary strengths, i.e. good angular resolution for optical and good radial resolution for \hi intensity maps. We therefore suggest potential methods to mitigate the effects which a blind foreground clean has on \hi intensity maps. These not only serve to drastically improve cross-correlations with photometric optical data, but also provide additional improvements in cross-correlations with spectroscopic galaxy surveys, as well as \hi intensity mapping auto-correlations. The two methods we propose are:

\begin{itemize}[leftmargin=*]
\item \bf{LoS Mean Reconstruction:} \normalfont{This is theoretically possible using optical galaxies which measure density along the LoS. By relating the optical over-density to the \hi temperature we can make a prediction for the LoS mean \hi temperature that has been removed and reverse the effect of this loss of information.}
\\
\item \bf{Artificial Extension to Redshift Range:} \normalfont{Introducing a buffer at either end of the data sets in the redshift (or frequency) direction will limit edge effects and as we will demonstrate, improves the general agreement with the original data.}
\end{itemize} 

\noindent We discuss both of these methods in more detail in the following sub-sections. 

\subsubsection{Line-of-Sight Mean Reconstruction}\label{LoSReconstructionSection}

While recovering the exact LoS means from the intensity map data is not possible (they are inaccessible before the clean, and removed after it) we can make predictions of what they are from  other data. Then by measuring the power spectrum of the LoS mean predictions, we can reverse the effects of the mean removal. Making use of our hypothesis in Equation \eqref{LoSReconstructionEq} we can write
\begin{equation}\label{crosscorrection}
    \langle \delta_\text{g} \delta T_\hiindex \rangle = \langle \delta_\text{g} \delta T_\text{clean} \rangle + \langle \delta_\text{g} \overline{\delta T}_\text{LoS} \rangle \, ,
\end{equation}
and similarly for the auto-correlation we have
\begin{equation}\label{autocorrection}
    \langle \delta T_\hiindex \delta T_\hiindex \rangle = \langle \delta T_\text{clean} \delta T_\text{clean} \rangle + 2\langle \delta T_\text{clean} \overline{\delta T}_\text{LoS} \rangle + \langle \overline{\delta T}_\text{LoS} \overline{\delta T}_\text{LoS}  \rangle \, .
\end{equation}
Therefore, for a cross-correlation we require the correction term $\langle \delta_\text{g} \overline{\delta T}_\text{LoS} \rangle$ and for an auto-correlation we require $2\langle \delta T_\text{clean} \overline{\delta T}_\text{LoS} \rangle + \langle \overline{\delta T}_\text{LoS} \overline{\delta T}_\text{LoS}  \rangle$. We can utilise the optical number density fields to make estimates for these terms. This is because we can relate the optical over-density $\delta_\text{g} = b_\text{g}\delta_\text{M}$ to temperature fluctuations $\delta T_\hiindex = \overline{T}_\hiindex b_\hiindex \delta_\text{M}$ through 
\begin{equation}
    \delta T_\text{orig}(z_i) = \frac{\overline{T}_\hiindex(z_i) b_\hiindex(z_i)}{b_\text{g}(z_i)} \delta_\text{g}(z_i) \, \, .
\end{equation}
Then we relate this to each LoS mean by
\begin{equation}
    \overline{\delta T}_\text{LoS} = \frac{1}{N_z} \sum_i \frac{\overline{T}_\hiindex(z_i) b_\hiindex(z_i)}{b_\text{g}(z_i)} \delta_\text{g}(z_i).
\end{equation}
This is all that is required to construct the correction terms for the cross- and auto-correlations outlined by Equations \eqref{crosscorrection} and \eqref{autocorrection}. This approach does not require precise optical redshift information for the $\delta_\text{g}(z)$. It is sufficient to use the poorly constrained photometric redshifts since the error on these should not heavily impact on the slowly varying summation kernel $\overline{T}_\hiindex(z) b_\hiindex(z)/b_\text{g}(z)$.

The prefactor $\overline{T}_\hiindex(z) b_\hiindex(z)/b_\text{g}(z)$ is not directly observable and therefore requires independent modelling or indirect measurement. $\overline{T}_\hiindex$ (Equation \eqref{TbarModelEq} and discussed thereafter) is degenerate with $b_\hiindex$. Note that redshift space distortions can break this degeneracy and constrain $\Omega_\hiindex$ and consequently $\overline{T}_\hiindex$ \citep{GBTHIdetection1}. For the purpose of testing this correction method we assume $\overline{T}_\hiindex$ has been accurately obtained, i.e. we simply use the model \eqref{TbarModelEq} which our simulated intensity maps have been designed to conform to. For the bias terms we determine them based on fiducial models where
\begin{equation}\label{gbiasEq}
	b_\text{g}(z) = 1 + 0.84z \, ,
\end{equation}
which was estimated from simulation results in \citet{Weinberg:2002rm} and used in the LSST Science book \citep{Abell:2009aa}. Following \citet{Bacon:2018dui} we model the \hi bias as
\begin{equation}\label{HIbiasEq}
	b_\hiindex(z) = 0.67 + 0.18z + 0.05z^2 \, . 
\end{equation}

\subsubsection{Artificial Extension to Redshift Range}\label{zExtSection}

\begin{figure}
  	\includegraphics[width=\columnwidth]{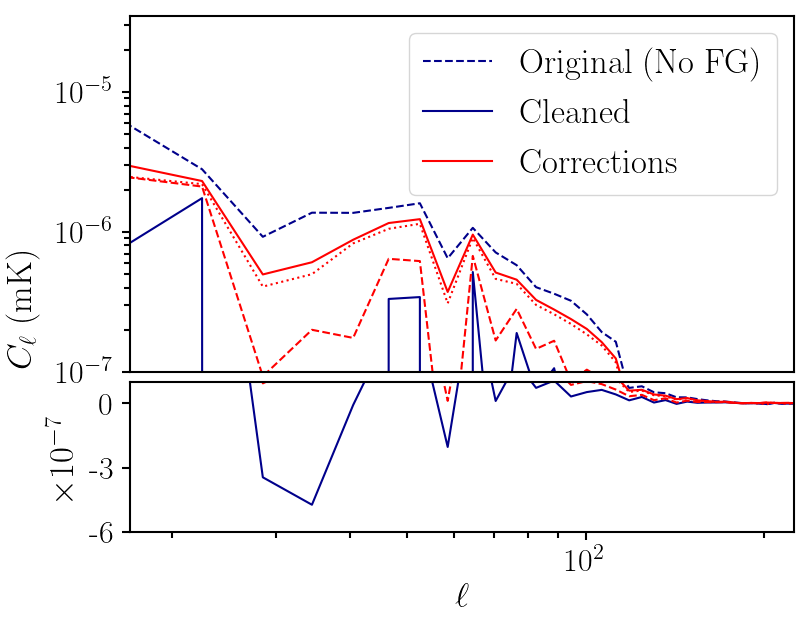}
    \caption{Cross-correlation angular power spectrum for the MICE simulation at redshift of $z=1.075$ and like Figure \ref{DES-photozCl} we have used the DES-like photometric redshifts available in MICE. Again the impact from foregrounds is visible in the difference between the blue dashed line and blue solid line. However, the effectiveness of the corrective techniques that we outlined in Section \ref{CorrectionsSection}, shown by the red lines, is encouraging. The dashed red line is for the LoS mean correction, the dotted red line represents the extended-$z$ correction and the red solid line represents both corrections applied. Produced using bandpowers with 6 multipoles per bin for clarity.}
    \label{Cl-DES_corrected}
\end{figure}

While the reconstruction of the LoS means works reasonably well for the mid-range redshifts, improvements can still be made especially to the edge effects caused by a foreground clean. These edge effects have been previously noted and suggestions have been made to exclude these contaminated regions \citet{Wolz:2013wna,Wolz:2015sqa}. One simple solution to mitigate this effect and limit the data excluded, is to extend the range of the data with the idea that the new artificial edges suffer the edge effect problems, but can then be removed from the rest of the data. We therefore take the full observed signal in the original $N$ redshift bins given by
\begin{equation*}
    [z_1,z_2,...,z_{N-1},z_N]
\end{equation*}
and pad both ends with replicated reversed data to become
\begin{equation*}
    [z_N,z_{N-1},...,z_2,z_1,z_1,z_2,...,z_{N-1},z_N,z_N,z_{N-1},...,z_2,z_1].
\end{equation*}
So we have added reversed copies of the data to the beginning and the end of the original redshift range. This ensures the padded data includes continuous foregrounds since this is what a blind foreground clean needs to utilise in order to remove them.\\
\newline
\noindent Figure \ref{Cl-DES_corrected} shows the performance of these corrections on the MICE catalogue. We have shown this at a redshift of $z=1.075$ which is closer to the extreme end of the redshift range for MICE and therefore has more need for correction. The solid blue line which shows the cross-correlation signal for \fastica foreground cleaned map demonstrates how poor the signal is without any correction. The solid red line then shows that with the artificial extension to the redshift ranges and the LoS mean corrections to the power spectrum outlined by Equation \eqref{crosscorrection}, the signal is significantly recovered and approaches the original signal with no foregrounds (blue dashed line).

\begin{figure}
  	\includegraphics[width=\linewidth]{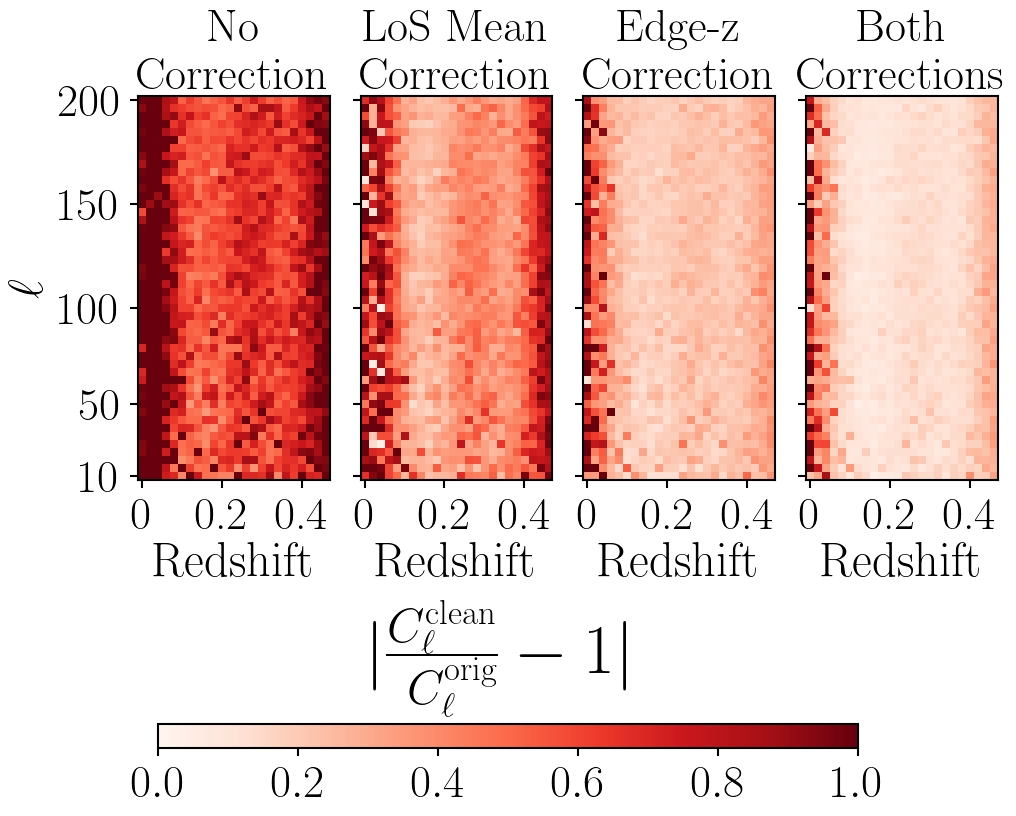}
    \caption{Demonstration of improvement on cross-correlation by including the corrections to the data outlined in Section \ref{CorrectionsSection}. This is for the GAEA data-set and shows relative differences for cross-correlation of optical photometric-like data with \hi intensity maps for the original (no foregrounds) and cleaned cases. For the optical sample we used a catalogue with redshift error of $\sigma_z = 0.06$.}
    \label{CorrectionComparison}
\end{figure}

We also demonstrate the more general improvement made across all redshift bins with Figure \ref{CorrectionComparison} which is for the GAEA simulation. Using the relative difference between original and clean power spectra as a gauge of performance (stated above the colour-bar), this shows how improved the signal is across all redshifts and scales with the corrections in place. We still see some poor disagreement in the very first redshift bin and slightly poorer performance for the last few bins, but the catastrophic discrepancies that we were seeing previously have been addressed.

These results are encouraging and suggest that with further refinement and understanding, cross-correlations between foreground cleaned intensity maps and photometric imaging surveys should be a useful probe of cosmology. We stress that our suggested corrections need further testing, preferably alongside real data to ensure they are reliable. 

\section{Clustering-Based Redshift Estimation}\label{ClusteringzSec}

As a direct example of the potential impact that foreground removal can have on cross-correlations with photometric redshift surveys, we now aim to use our simulations to see if a photometric calibration method using such cross-correlations is still viable. This method utilises the shared clustering signal between photometric optical galaxies and overlapping \hi intensity maps. This clustering-based redshift estimation process has previously been studied in \citet{AlonsoIMClusteringz} and \citet{Cunnington:2018zxg}, but a full analysis including simulated foreground contamination has not yet been conducted.

Given the difficulties outlined in Section \ref{HIxOpticalSec}, such a method represents a stern test since the intensity maps are correlated with a population of optical galaxies where little redshift information is assumed. The only assumption made is that the optical galaxies are within the redshift range covered by the reference intensity maps. This is  applicable to weak-lensing probes where wide redshift bins may be used, and where the aim is to obtain the source distribution which is required for precise measurements of cosmological shear. This wide redshift binning would mean huge degradation in small-scale radial modes, which is a major obstacle for this method given the increased noise due to the redshift uncertainty as outlined in Figure \ref{PhotozCls}.

\subsection{HI Clustering-$z$ Method}\label{ClusteringzMethod}

In order to make a prediction for the redshift distribution of optical galaxies we require an estimator which utilises the shared clustering signal between the opticals and the \hi intensity maps. We use the following estimator and refer the reader to \citet{Cunnington:2018zxg} where a full derivation is given:
\begin{equation}\label{dNdzEstimator}
    \frac{\text{d}N_\text{g}}{\text{d}z}(z) = \frac{w_\text{g,\hiindex}(z)}{w_\text{\hiindex,\hiindex}(z)}\overline{T}_\hiindex(z)\frac{b_\hiindex(z)}{b_\text{g}(z)}\frac{1}{\Delta z} \, .
\end{equation}
Here we use angular correlations functions $w$ where $w_\text{g,\hiindex}(z)$ is the cross-correlation between all the optical galaxies and an \hi intensity map at a redshift $z$. Similarly, $w_\text{\hiindex,\hiindex}(z)$ is the auto-correlation between two intensity maps at redshift $z$.  An effective test of this estimator given the contamination of foregrounds is to use information from the $C_\ell$ power spectra since this is a measurement of angular clustering which is what we want to utilise for estimating $\text{d}N_\text{g}/\text{d}z$. An effective measurement for the angular correlation functions, which closely follows previous clustering redshift work \citep{MenardClustering-z} is given by
\begin{equation}
    w_{XY}(z) = \int_{\ell_\text{min}}^{\ell_\text{max}} W(\ell) C^{XY}_\ell(z)  \text{d}\ell \, ,
\end{equation}
where $W(\ell)$ is a weight function which can be tuned to certain scales. For our purposes $W(\ell) = \ell$ is sufficient to give weight to smaller scales where more useful matching is expected to exist. As previously, the indexes $X$ and $Y$ can either be chosen to represent the \hi intensity map auto-correlation where $X=Y=\hiindex$ or the cross-correlation with the optical where $X=\text{g}$ and $Y=\hiindex$.

As before, $\overline{T}_\hiindex$ is the average brightness temperature which is known in our simulations. In reality however, the observable is a temperature fluctuation and $\overline{T}_\hiindex$ requires modelling as explained previously in equation \eqref{TbarModelEq}. Again, for our purposes we assume an accurate modelling of $\overline{T}_\hiindex$ has been achieved, i.e. we simply measure the quantity in our simulations.

\begin{figure*}
    \includegraphics[width=1\linewidth]{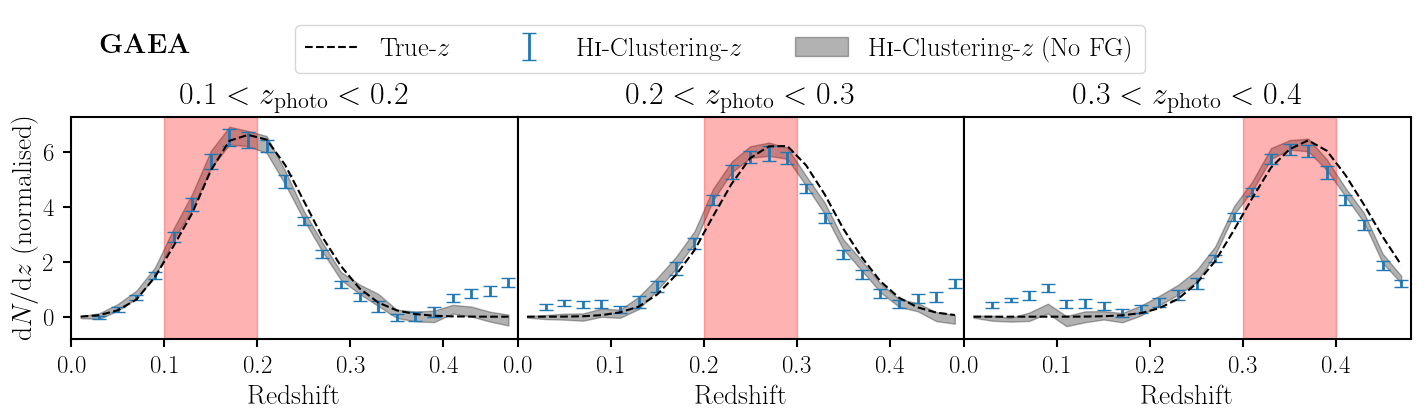}
\end{figure*}
\begin{figure*}
    \includegraphics[width=1\linewidth]{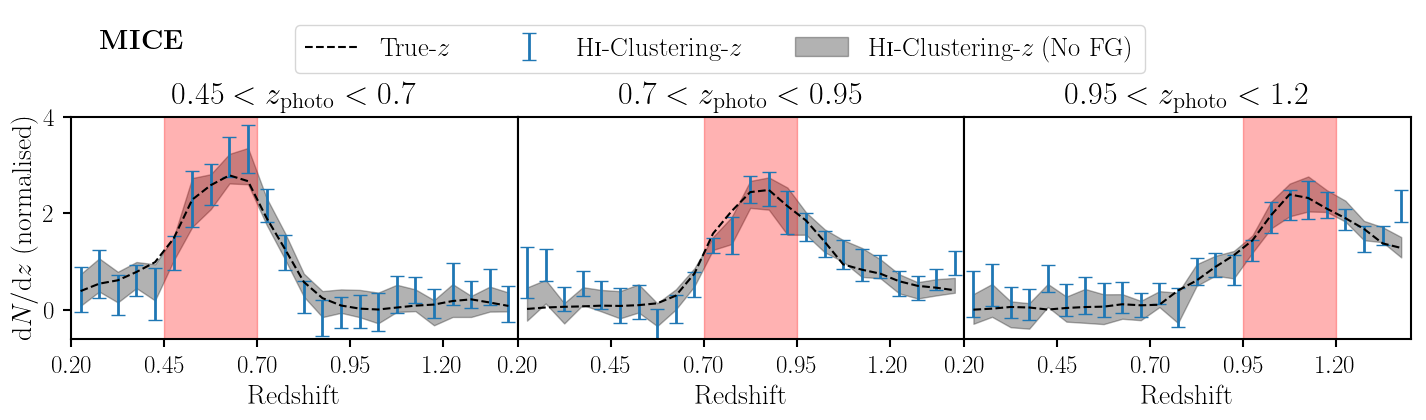}
\caption{Clustering-based redshift estimation results using both the GAEA and MICE catalogues. 
The pink vertical shaded regions represent the optical sample chosen as galaxies whose photometric redshift lies within the targeted redshift ranges. 
The black dashed lines show the true redshift distributions of these galaxies.
The blue data points give the estimated redshift distributions based on cross-correlations with \hi intensity maps and using the estimator in Equation \eqref{dNdzEstimator}. Intensity maps have foregrounds added and then removed with the \fastica process with corrections made (Section \ref{CorrectionsSection}). We also include the estimated distributions with errors from intensity maps absent from any foreground contamination, shown as the grey shaded distribution. The GAEA model uses a beam size of $\theta_\text{beam} = 1.46^\text{o}$, representative of an SKA-like beam for that redshift range. However, for MICE we have used a smaller beam size of $\theta_\text{beam} = 1^\text{o}$ because of its smaller sky coverage.}
\label{dNdzphotoz}
\end{figure*}

Finally, the estimator in Equation (\ref{dNdzEstimator}) also requires the bias ratio $b_\hiindex/b_\text{g}$. We can find this from the angular auto-correlation power spectra for the two samples:
\begin{equation}
	\frac{b_\hiindex(z)}{b_\text{g}(z)} = \frac{1}{\overline{T}_\hiindex(z)}\sqrt{\frac{C_{\hiindex\hiindex}(\ell,z)}{C_\text{gg}(\ell,z)}} \, .
\end{equation}
However this relies on binning the galaxies by true redshift to measure the bias at that redshift. But we choose to assume that the optical sample has very poorly known (effectively unconstrained) redshifts, since it will be these surveys where redshift calibration is most in demand, so obtaining $C_\text{gg}(z)$ accurately is not possible. For this study we therefore rely on fiducial models of the individual biases as laid out in Equations \eqref{gbiasEq} and \eqref{HIbiasEq}.

\subsection{HI Clustering-$z$ Results}

We are now ready to present a simple test of our \hi clustering-based redshift estimation method and demonstrate its capability of recovering a redshift distribution using the \hi intensity maps discussed in Section \ref{HIIMSec} for our simulated optical photometric sample with a detection threshold applied as discussed in Section \ref{PhotozcatSec}. This is all in the presence of 21cm foreground contamination which has been cleaned using a \fastica process. We also apply our corrections as outlined in Section \ref{CorrectionsSection} using only the photometric redshift information available.

We test this approach on both our GAEA and MICE based simulations and Figure \ref{dNdzphotoz} shows the results. In both cases we select optical galaxies based on their photometric redshifts in targeted redshift ranges shown as the pink shaded regions on the plots. Because these galaxies have been selected using their poorly constrained photometric redshifts, the true redshift distribution (black dashed line) extends way beyond these ranges. By cross-correlating with \hi intensity maps and using the estimator outlined by Equation \eqref{dNdzEstimator} we can make a prediction of this true redshift distribution, shown by the blue data points. The grey shaded distributions show the result without any foreground contamination. We obtain the error bars for $\text{d}N/\text{d}z$ using a jackknifing technique, gridding the maps into an array of 25 smaller sub-samples. We then measure our estimator on the map but omit one of the 25 sub-samples. We repeat the procedure, averaging over the estimators obtained from omitting sub-samples, and obtain a standard deviation.

These results are very encouraging for the future of using shared clustering signals from \hi intensity maps to calibrate photometric redshifts. A small bias is present which appears to skew the distribution, most evident in the GAEA results where the error is low. This will be caused by the fiducial bias models we use (Equations \eqref{gbiasEq} and \eqref{HIbiasEq}) in the estimator \eqref{dNdzEstimator} not agreeing precisely with the simulated catalogues. More focused follow-up on this bias factor is required, as discussed in the previous section, and an improved approach which constrains the biases and mean \hi temperature should mitigate this slight skewness.

Small discrepancies tend to exist at the extreme ends of the redshift distribution. When the true redshift distribution at these edges should be close to zero, often the estimator in the foreground contaminated case, predicts a non-zero quantity. These are due to residual edge effects not fully mitigated by the correction outlined in Section \ref{zExtSection}. Because of this, it is difficult to place quantitative interpretation on the results in Figure \ref{dNdzphotoz} without these edge discrepancies skewing the measurement. We calculate the Median Absolute Deviation (MAD) for the differences between the true and estimated distributions for the foreground contaminated blue data points2 i.e. $\text{d}N/\text{d}z_\text{true} - \text{d}N/\text{d}z_\text{est}$, since this measurement will not be too sensitive to the incorrect estimations near the edges. We find that for the three GAEA distributions shown in Figure \ref{dNdzphotoz} these MAD values are 0.199, 0.167 and 0.284 for $0.1<z_\text{photo}<0.2$, $0.2<z_\text{photo}<0.3$ and $0.3<z_\text{photo}<0.4$ respectively. For MICE, the MAD values for the differences in true and estimated distributions are 0.129, 0.107 and 0.132. The similar values in each simulation demonstrate that the redshift prediction method is behaving consistently. The relatively low MAD values, under 5\% of the normalised $\text{d}N/\text{d}z$ peak value, also suggest the discrepancies between true and estimated distributions are mostly small and is indicative of the estimator's precision. This represents an excellent test of cross-correlations between foreground affected \hi intensity maps and photometric surveys. This is because this method relies on sufficient cross-signal existing for poorly constrained optical redshifts over wide redshift ranges. The relative success of this method suggests that the problems outlined in Section \ref{PhotozUncertainty} will be surmountable.

We found that a key factor regarding the success of the clustering-based redshift estimation method using \hi intensity maps is the combination of the sky area and the size of the instrumental beam. \citet{Cunnington:2018zxg} found that the error on the estimation is directly proportional to the beam size and can be approximated by 
\begin{equation}\label{sigsmootherror}
    \sigma_{N(z)} \propto \frac{\theta_\text{beam}}{\sqrt{A}} \, ,
\end{equation}
where $A$ is the area of the sky covered. Due to the smaller sky coverage in the MICE simulation we found that we were unable to use a constant beam size of $\theta_\text{beam}=2.36^\text{o}$ which would be representative of an SKA-like beam probing redshifts up to $z=1.4$. Instead we have only smoothed with a $1^\text{o}$ beam. However, having larger sky coverage in future simulations  would mitigate this issue. It is interesting to note how the error does not increase too much in Figure \ref{dNdzphotoz} with the inclusion of foreground contamination in the analysis (comparison between blue data points and grey shaded distribution). This supports the claim that the error from this estimator is largely dominated by the sky area and beam size and explains the larger errors on the MICE plot compared with GAEA.

\section{Summary \& Conclusions}\label{conclusion}

Forthcoming \hi intensity mapping experiments will be able to contribute to cosmological studies through \hi auto-correlations as well as cross-correlations with optical galaxy surveys. To ensure that \hi intensity mapping is a competitive technique, it is important to understand 21cm foreground contamination, and the effects of foreground removal on the measurements.

In this work we have taken a simulations-based approach to investigate these issues, focusing on the foreground removal effects on \hi intensity mapping cross-correlations with photometric galaxy surveys. By using existing $N$-body simulations and the galaxy catalogues produced from them, we constructed both optical galaxy catalogue data and \hi intensity map data with the same underlying cosmological clustering signal. We then simulated the relevant 21cm foreground signals that are expected to contaminate the \hi intensity maps, and used a state-of-the-art blind foreground removal process known as \fastica. This approach allowed us to then examine what impact this type of foreground removal has on cosmological probes such as the clustering measured by the angular power spectrum $C_\ell$.\newline
\\
\noindent Our main conclusions are as follows:

\begin{itemize}[leftmargin=*]

\item We have shown evidence that a \fastica reconstruction will successfully allow accurate auto-correlation measurements as shown by previous work \citep{Wolz:2013wna, Shaw:2013wza}. Figure \ref{Clauto_NIC} showcases the results for both our simulations, GAEA and MICE. The better result obtained for the GAEA model is likely due to its larger sky size allowing for more samples to average over in negentropy calculations. 
\\
\item The auto-correlation tests we performed strongly suggest that a frequency dependent beam size will cause problems for independent component-like methods as demonstrated in Figure \ref{Clauto_NIC} and also shown by \citet{Alonso:2014dhk}. A solution to this is to re-smooth the intensity maps to match the beam size for the highest redshift when using these foreground removal techniques.
\\
\item \fastica also delivers good results in cross-correlation with optical galaxy data where the redshifts for the opticals are very well constrained as they would be in a spectroscopic-like survey. In Figure \ref{Cl_cross-zbinned} we used optical galaxies with true redshifts in cross-correlation with \hi intensity maps. The figure shows the excellent agreement between using the original (no foregrounds included) intensity maps and the foreground cleaned ones.
\\
\item We find that further treatment is needed when cross-correlating foreground cleaned \hi intensity maps with photometric-like optical galaxy surveys with poor redshift constraints. Figures \ref{PhotozCls} and \ref{DES-photozCl} show the impact of combining foreground cleaned intensity maps with an imaging galaxy survey which has poorly constrained redshifts. This poor result is unsurprising and can be generally explained by the combination of eroded large-radial modes caused by the foreground cleaning, with eroded small radial-modes caused by the uncertainty in the photometric redshifts \citet{Witzemann:2018cdx}.
\\
\item More specifically, we find that a cause of the poor results when considering \hi $\times$ Photo-$z$ is the loss of LoS mean information when conducting the foreground clean. Figure \ref{LosMeans} shows how any prior off-zero LoS means are collapsed to zero which has the effect of unpredictably changing pixel values in the transverse maps, as demonstrated in Figure \ref{LoSReconstruction}. As a possible treatment for this unwanted effect we proposed a LoS reconstruction that uses information from the optical galaxies as outlined by Equation \eqref{LoSReconstructionEq}. This, coupled with artificially extending the redshift range to mitigate the edge effects caused by the foreground clean, improves results as shown by Figures \ref{Cl-DES_corrected} and \ref{CorrectionComparison}.
\\
\item Finally, we conducted a comprehensive test of these methods by attempting to use foreground contaminated intensity maps for clustering-based redshift estimation of a photometric optical sample. By using \fastica and our additional corrections we were able to accurately predict the redshift distributions for mock optical catalogues in both our models (Figure \ref{dNdzphotoz}).

\end{itemize} 

\noindent This work used two independent $N$-body simulations, where one (GAEA) used a semi-analytical approach to constructing a galaxy catalogue and the other (MICE) used a HOD/HAM hybrid method. The resulting catalogues formed the basis for constructing the optical and \hi intensity map mock data. This means we can be confident that the conclusions we have made are unlikely to be specific to these simulations. 

A limitation in using existing mock galaxy catalogues to generate \hi intensity maps however comes from the finite number of galaxies available to sample in the map. The great advantage of \hi intensity mapping is the frequency resolution which allows for numerous tomographic bins. While the catalogues we use are large (>$10^8$ galaxies), this finite number means care was needed when going to large numbers of tomographic bins. If the bin is too thin, it will contain a low number of galaxies (sparse galaxy density), and therefore a sparse signal in each pixel. This is not an accurate emulation of an intensity map which should provide a near continuous emission profile. Tests were carried out with a higher number of bins in some cases. For example we used 240 redshift bins for the MICE catalogue and tested if we still see the LoS mean destruction demonstrated by Figure \ref{LosMeans}. Indeed we find that even with this more realistic number of bins, we find similar results but cannot be certain that these are accurate simulations of combined emission maps since the number density of simulated galaxies becomes low ($\sim 5$ per voxel) at this fine radial resolution. This is why we used relatively thick tomographic bins in this work ($\Delta z = 0.02$ for GAEA and $\Delta z = 0.05$ for MICE). Furthermore, it is likely that intensity maps would need to be integrated to this size of bin when using cross-correlations with photometric surveys to measure power spectra or measure cosmological parameters. This is because the redshift uncertainty on the opticals would demand a thick tomographic bin to ensure enough signal-to-noise.

Throughout this work we have made assumptions that parameters such as the mean \hi temperature ($\overline{T}_\hiindex$) can be precisely obtained. While we use a model for this parameter in our analysis, this same model was used in the construction of the \hi intensity map signal, therefore its success is unsurprising. However, other parameters such as the clustering bias terms ($b_\text{g}$ and $b_\hiindex$) are not directly fed into our simulated signals, so the success of modelling these as scale-independent biases in our clustering-based redshift estimation is encouraging.

Note that in this work we have not simulated any foreground polarization leakage effects. However, in many frequency channels we have smoothed our maps more than is required to simulate the instrument beam, which is a treatment previously used in real data to mitigate these effects \citep{GBTHIdetection2}. It is unclear whether the required level of instrument calibration is achievable to avoid effects such as polarization leakage. Therefore one could argue that it will not necessarily be the foregrounds themselves that cause the biggest problems, but instead the leakage of them through imperfect instrument calibration \citep{Moore:2013ip, Shaw:2014khi}. Therefore, a follow-up study with simulations of realistic observations including polarization leakage and other instrument systematics such as $1/f$ gain fluctuations, beam side-lobes, radio-frequency interference etc. \citep{Harper:2018dys} will be an important step.

Furthermore, in this work we did not consider the clustering of point source foregrounds, which one could argue has potential to bias cosmological clustering measurements. Nor in our simulations did we simulate the anisotropy of galactic free-free emission which is expected to be stronger in the galactic plane. However, neither of these subtle features are likely to affect the frequency coherence of the signals which \fastica uses to isolate them.

In future work we plan to include a further analysis into the effects of foreground removal on  cosmological measurements including the 3D correlation function $\xi(s)$ and power spectrum $P(k)$ multipoles, extending the work of \citet{Blake:2019ddd}.

As \hi intensity mapping data becomes available alongside the plethora of high precision optical datasets, we will be able to confirm conclusions derived from simulated mocks using real observations. Future measurements of \hi $\times$ Photo-$z$ data, for example from MeerKAT and DES \citep{Pourtsidou:2017era, Bacon:2018dui} or TIANLAI \citep{Tianlai} and DECaLS\footnote{http://legacysurvey.org/} \citep{DECALS}, will be an excellent test for our claims in this paper. We have demonstrated the potential of such experiments with our example of how cross-correlations can be used for photometric redshift calibration. This is a major challenge for forthcoming Stage-IV instruments utilising photometric optical samples, such as LSST and \euclid.
We believe that photometric redshift calibration using \hi intensity mapping data is an alternative method with great promise for tackling this challenge.

To summarise, we have shown evidence that a method such as \fastica performs excellently at reconstructing the inherently weak \hi signal in the presence of dominant 21cm foreground contamination. Even in cross-correlation with optical data with poorly constrained redshifts, with our suggested corrections it is possible to make good measurements of the cosmological signal. We have introduced a LoS mean reconstruction as a treatment for foreground cleaned intensity mapping signal loss, which improves the fidelity of cross-correlation measurements but which will benefit from further investigation and refinement.
Foreground contamination is a challenge for \hi intensity mapping, but this work alongside others demonstrates that it is a surmountable one. We look forward to providing even more realistic simulations, and testing our proposed methods with real data, in the near future.

\section*{Acknowledgements}

We are grateful to Chris Blake, Emma Chapman, Pablo Fosalba, Ian Harrison, Mario Santos and Anna Zoldan for many useful discussions and feedback. We also
extend our gratitude to the referee whose report improved
the quality of this paper. 

SC is supported by the University of Portsmouth. AP is a UKRI Future Leaders Fellow and has also received support from STFC Grant ST/S000437/1. DB is supported by STFC Grant ST/N000668/1. 

This work has made use of CosmoHub. CosmoHub has been developed by the Port d'Informacio Cientifica (PIC), maintained through a collaboration of the Institut de Fisica d'Altes Energies (IFAE) and the Centro de Investigaciones Energeticas, Medioambientales y Tecnologicas (CIEMAT), and was partially funded by the "Plan Estatal de Investigacion Cientifica y Tecnica y de Innovacion" program of the Spanish government \citep{Carretero:2017zkw}.

Numerical computations for this research were done on the Sciama High Performance Compute (HPC) cluster which is supported by the ICG, SEPNet, and the University of Portsmouth.




\bibliographystyle{mnras}
\bibliography{Bib} 


\bsp	
\label{lastpage}
\end{document}